\documentclass[aps,prl,preprintnumbers,showpacs,nofootinbib,floatfix]{revtex4}
\usepackage{graphicx}
\usepackage{subfig}
\usepackage{comment}
\usepackage{dcolumn}
\usepackage{bm}
\begin{document}
\newcommand{\newc}{\newcommand}
\newc{\ra}{\rightarrow}
\newc{\lra}{\leftrightarrow}
\newc{\beq}{\begin{equation}}
\newc{\eeq}{\end{equation}}
\newc{\barr}{\begin{eqnarray}}
\newc{\earr}{\end{eqnarray}}
\newcommand{\Od}{{\cal O}}
\newcommand{\lsim}   {\mathrel{\mathop{\kern 0pt \rlap
  {\raise.2ex\hbox{$<$}}}
  \lower.9ex\hbox{\kern-.190em $\sim$}}}
\newcommand{\gsim}   {\mathrel{\mathop{\kern 0pt \rlap
  {\raise.2ex\hbox{$>$}}}
  \lower.9ex\hbox{\kern-.190em $\sim$}}}
  \def\rpm{R_p \hspace{-0.8em}/\;\:}
\title {HALO DENSITY PROFILES CONSISTENT WITH ASYMMETRIC M-B VELOCITY DISTRIBUTIONS-IMPLICATIONS ON DIRECT DARK MATTER SEARCHES}
\author{ J.D. Vergados}
\affiliation{University of Ioannina, Ioannina, GR 45110, Greece \footnote {e-mail: vergados@cc.uoi.gr}}
\vspace{0.5cm}
 \begin{abstract}

 In the
present paper  we obtain the WIMP velocity distribution in our vicinity starting from spherically symmetric WIMP density profiles in a self consistent way by employing the Eddington approach. By adding a reasonable angular momentum  dependent term in the expression of the energy, we obtain axially symmetric WIMP velocity distributions as well. We find that some density profiles lead to approximate Maxwell-Boltzmann distributions, which are  automatically defined in a finite domain, i.e. the escape velocity need not be put by hand. The role of such distributions in obtaining the direct WIMP detection rates, including the modulation,  is studied in some detail and, in particular, the role of the asymmetry is explored.
\end{abstract}
\maketitle

\section{Introduction}
The combined MAXIMA-1 \cite{MAXIMA1},\cite{MAXIMA2},\cite{MAXIMA3}, BOOMERANG \cite{BOOMERANG1},\cite{BOOMERANG2},
DASI \cite{DASI1} and COBE/DMR Cosmic Microwave Background (CMB)
observations \cite{COBE} imply that the Universe is flat
\cite{flat01}, $\Omega=1.11\pm0.07$ and that most of the matter in
the Universe is Dark \cite{SPERGEL}. i.e. exotic. Combining the
recent WMAP data \cite{WMAP08a} with other experiments one finds:
$$\Omega_b=0.0456 \pm 0.0015, \Omega _{CDM}=0.228 \pm 0.013 , \Omega_{\Lambda}= 0.726 \pm 0.015$$
Since the non exotic component cannot exceed $40\%$ of the CDM
~\cite {Benne}, there is room for the exotic WIMP's (Weakly
Interacting Massive Particles).
 Supersymmetry naturally provides candidates for the dark matter constituents
 \cite{ELLROSZ}-\cite{GOODWIT}.
 In the most favored scenario of supersymmetry the
LSP (Lightest Supersymmetric Particle) can be simply described as a Majorana fermion, a linear
combination of the neutral components of the gauginos and
higgsinos \cite{GOODWIT},\cite{ref2a},\cite{ref2b},\cite{ref2c},\cite{ref2d}. In most calculations the
neutralino is assumed to be primarily a gaugino, usually a bino.
 Other particle models have also been considered, like Kaluza-Klein WIMPs (see,e.g. the recent work \cite{OikVerMou} and references there in ), sterile neutrinos \cite{LaiShap08},  technicolor \cite{GudKouv06} and   recently  composite WIMP's \cite{KHLKOUV08} (see also  the recent theory review \cite{DUSEL1}).

 Even though there exists firm indirect evidence for a halo of dark matter
 in galaxies from the
 observed rotational curves, it is essential to directly
detect \cite{GOODWIT}-\cite{KVprd}
 such matter.
Until dark matter is actually detected, we shall not be able to
exclude the possibility that the rotation curves result from a
modification of the laws of nature as we currently view them.  This makes it imperative that we
invest a
maximum effort in attempting to detect dark matter whenever it is
possible. Furthermore such a direct detection will also
 unravel the nature of the constituents of dark matter.\\
 The
 possibility of such detection, however, depends on the nature of the dark matter
 constituents (WIMPs).
  Since the WIMP is expected to be very massive, $m_{\chi} \geq 30 GeV$, and
extremely non relativistic with average kinetic energy $T \approx
50KeV (m_{\chi}/ 100 GeV)$, it can be directly detected
~\cite{GOODWIT}-\cite{KVprd} mainly via the recoiling of a nucleus
(A,Z) in elastic scattering. The event rate for such a process can
be computed from the following ingredients:
\begin{enumerate}
\item An effective Lagrangian at the elementary particle (quark)
level obtained in the framework of supersymmetry as described ,
e.g., in Refs~\cite{ref2a}-\cite{ref2d},\cite{JDV96}.
\item A well defined procedure for transforming the amplitude
obtained using the previous effective Lagrangian from the quark to
the nucleon level, i.e. a quark model for the nucleon. This step
is not trivial, since the obtained results depend crucially on the
content of the nucleon in quarks other than u and d. This is
particularly true for the scalar couplings, which are proportional
to the quark masses~\cite{Dree1}$-$\cite{Chen2}, \cite{JDV06a} as well as the
isoscalar axial coupling \cite{JELLIS93a,JDV06a}.
\item Knowledge of the relevant nuclear matrix elements
\cite{Ressa},\cite{Ressb},\cite{DIVA00}, obtained with as reliable as possible many
body nuclear wave functions. Fortunately in the case of the scalar
coupling, which is viewed as the most important, the situation is
a bit simpler, since  then one  needs only the nuclear form
factor.
\item Knowledge of the WIMP density in our vicinity and its velocity distribution. Since the
essential input here comes from the rotational curves,  dark matter candidates other than the
LSP (neutralino) are also characterized
by similar parameters.
\end{enumerate}

In the past various velocity distributions have been considered.
The one most used is the isothermal Maxwell-Boltzmann velocity
distribution with $<\upsilon ^2>=(3/2)\upsilon_0^2$ where
$\upsilon_0$ is the velocity of the sun around the galaxy, i.e.
$220~km/s$. Extensions of this M-B distribution were also
considered, in particular those that were axially symmetric with
enhanced dispersion in the galactocentric direction
 \cite {Druka,Drukb,Verg00}.  In all such
distributions an upper cutoff $\upsilon_{esc}=2.84\upsilon_0$ was introduced
by hand, in the range obtained by Kochanek\cite{Kochanek95}. In a different approach  Tsallis type functions, derived from simulations of dark matter densities were employed, see e.g. recent calculations \cite{VerHanH} and references there in .

Non isothermal models have also been considered. Among those one should
mention the late infall of dark matter into the galaxy, i.e caustic rings
 \cite{SIKIVI1,SIKIVI2,Verg01,Green,Gelmini}, dark matter orbiting the
 Sun \cite{Krauss}, Sagittarius dark matter \cite{GREEN02}.

The correct approach in our view is to consider the Eddington
proposal \cite{EDDIN}, i.e. to obtain both the density and the
velocity distribution from a mass distribution, which depends both
on the velocity and the gravitational potential. Our motivation in
using Eddington \cite{EDDIN} approach to describing the density of
dark matter is found, of course, in his success in describing the
density of stars in globular clusters. Since this approach
adequately describes the distribution of stars in a globular
cluster in which the main interaction is gravitational and because
of its generality , we see no reason why such an approach should
not be applicable to dark matter that also interact
gravitationally.
It seems, therefore, not surprising that this  approach has been
used by Merritt \cite{MERRITT} and  applied to dark matter by
Ullio and Kamionkowski\cite{ULLIO} and more recently by us
\cite{OWVER},\cite{VEROW06}.

 It is the purpose of the present paper to extend the previous work and obtain a
dark matter velocity distribution, which need not be spherically symmetric,  consistent with
assumed halo matter distributions with a natural upper velocity
cut off. It will then be shown that this distribution can be approximated by a Maxwell-Boltzmann (M-B) distribution with a finite domain that depends on the asymmetry parameter. The distribution obtained will be used to calculate WIMP direct detection rates including the annual modulation, as a function of the asymmetry parameter.
\section{The Dark Matter Distribution in the Context of  the Eddington approach}
As we have seen in the introduction the matter distribution can be given\cite{VEROW06} as follows
\beq dM=2\pi~f(\Phi({\bf r}),\upsilon_r,\upsilon_t)~dx~dy~dz
~\upsilon_t~d\upsilon_t~d\upsilon_r \label{distr.1} \eeq where the
function $f$ the distribution function, which depends on ${\bf r}$
through the potential $\Phi({\bf r})$ and the tangential and
radial velocities $\upsilon_t$ and $\upsilon_r$. 
We will limit ourselves in spherically symmetric systems.
 Then the density of matter $\rho(|r|)$
satisfies the equation:
  \beq
d\rho=2\pi~f(\Phi(|{\bf r}|
),\upsilon_r,\upsilon_t)~~\upsilon_t~d\upsilon_t~d\upsilon_r
\label{distr.2} \eeq
The distribution is a function of the total energy:
\begin{itemize}
\item The energy E is $\Phi(r)+\frac{\upsilon^2}{2}$. Then
\beq
\rho(r)=4 \pi\int f(\Phi(r)+\frac{\upsilon^2}{2})\upsilon^2 d\upsilon=4 \pi\int_{\Phi}^0 {f(E)}{\sqrt{2(E-\Phi)}}dE
\label{Eq:rhoE}
\eeq
This is an integral equation of the Abel type. It can be inverted to yield:
\begin{equation}
f(E)=\frac{\sqrt{2}}{4 \pi^2}\frac{d}{dE} \int_E^0 \frac{d \Phi}{\sqrt{\Phi-E}}\frac{d \rho}{d \Phi}
\end{equation}
The above equation can be rewritten as:
\begin{equation}
f(E)=\frac{1}{2 \sqrt{2} \pi ^2} \left[ \int_E^0 \frac{d\Phi}{\sqrt{\Phi-E}} \frac{d^2 \rho}{d \Phi ^2}-
\frac{1}{\sqrt{-E}} \frac{d \rho}{d \Phi}|_{\Phi=0} \right ]
\end{equation}
 In order to proceed it is necessary to know the density as a function of the potential. In practice only in few cases this can be done analytically. This, however, is not a problem, since this function can be given parametrically by the set $(\rho(r),\Phi(r))$ with the position $r$ as a parameter. The potential $\Phi(r)$ for a given density $\rho(r)$ is obtained by solving Poisson's equation.\\
Once the function f(E) is known we can obtain the needed velocity distribution $f_{r_s}(\upsilon)$ in our vicinity ($r=r_s$) by writing:
\beq
f_{r_s}(\upsilon)=f(\Phi(r)|_{r=r_s}+\frac{\upsilon^2}{2})/\rho(r=r_s)
\eeq
\item We suppose now that there is an additional kinetic term associated with an angular momentum \cite{bookBT87}, i.e. models of the Opsikov-Merritt type \cite{OPSIKOV79},\cite{MERRITT},\cite{MERRITT85b}  :
\beq
Q\equiv E+\frac{J^2}{2 r_0^2}=\Phi(r)+\frac{\upsilon^2}{2}+\frac{|r\times v|^2}{2 r^2_0}=\Phi(r)+\frac{\upsilon^2}{2}+\frac{r^2}{2r^2_0}\upsilon^2_t=
\Phi(r)+\frac{\upsilon^2_r}{2}+\left (1+\frac{r^2}{r^2_0}\right )\frac{\upsilon^2_t}{2}
\eeq
where $\upsilon_r$ and $\upsilon_t$ are the radial, i.e. outwards from the center of the galaxy, and the tangential components of the velocity
 and $r_0$ is the "anisotropy radius" to be treated as a phenomenological parameter. We now have:
  \beq
\rho=2\pi~\int~f\left ( \Phi({\bf r})
+\frac{\upsilon^2_r}{2}+\left(1+\frac{r^2}{r_0^2}\right)\frac{\upsilon^2_t}{2} \right )~~\upsilon_t~d\upsilon_t~d\upsilon_r
\label{distr.3} \eeq
The last integral takes the form:
\beq
\rho(r)=4 \pi \left(1+\frac{r^2}{r_0^2} \right )^{-1} \int_{\Phi}^0 {f(E)}{\sqrt{2(E-\Phi)}}dE
\eeq
This equation is formally  the same with Eq. (\ref{Eq:rhoE}) with the understanding that
\beq
\rho(\Phi)\rightarrow\tilde{\rho}(\Phi,r_0)=\rho(\Phi)\left (1+\frac{\left(r(\phi)\right)^2}{r_0^2} \right )
\eeq
The $r(\Phi)$ can be obtained by inverting the equation $\Phi=\Phi(r)$. In practice this is not needed, if, as we have mentioned above,
we use $r$ as a parameter.
We thus find:
\begin{equation}
f(E)=\frac{1}{2 \sqrt{2} \pi ^2} \left[ \int_E^0 \frac{d\Phi}{\sqrt{\Phi-E}} \frac{d^2 \tilde{\rho}}{d \Phi ^2}-
\frac{1}{\sqrt{-E}} \frac{d \tilde{\rho}}{d \Phi}|_{\Phi=0} \right ]
\end{equation}
The velocity distribution in our vicinity becomes
\beq
f_{r_s}(\upsilon)=f(\Phi(r)|_{r=r_s}+\frac{\upsilon^2_r}{2}+(1+\frac{r_s^2}{r_0^2})\frac{\upsilon^2_t}{2})
\eeq
and is only axially symmetric. The isotropic case follows as a special case in the limit $r_0\rightarrow0$.
\end{itemize}
 The characteristic feature of this approach is that the velocity distribution vanishes outside a given region specified by a cut off velocity $v_m$, by the positive root of the equation $f_{r_s}(\upsilon)=0$
\section{ Simple Realistic Dark Matter Density Profiles}
\begin{figure}[!ht]
 \begin{center}
 \subfloat
{
\rotatebox{90}{\hspace{-0.0cm}{$\frac{\rho(x)}{\rho_0}\longrightarrow$}}
\includegraphics[scale=0.6]{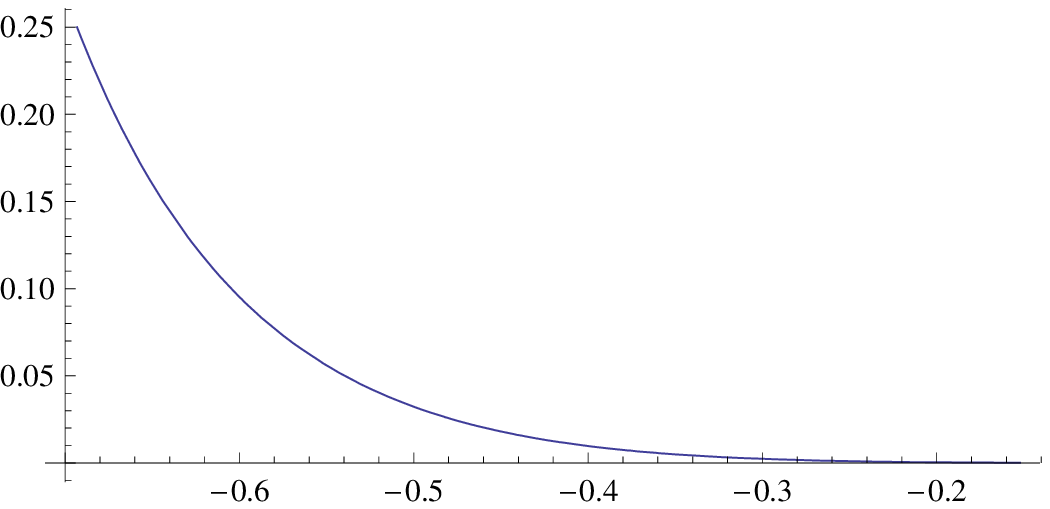}
}
 \subfloat
{
\rotatebox{90}{\hspace{-0.0cm}{$\frac{\rho(x)}{\rho_0}\longrightarrow$}}
\includegraphics[scale=0.6]{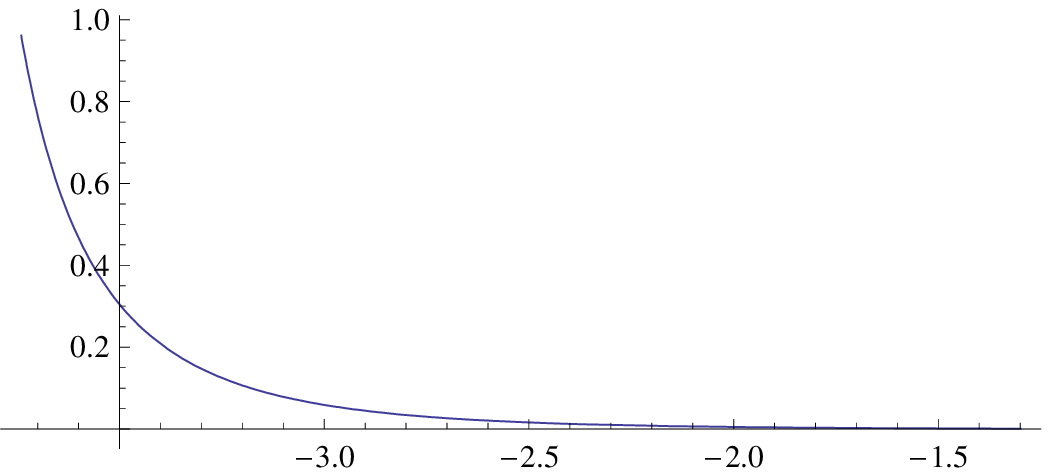}
}\\
{\hspace{-0.0cm} {$\frac{\Phi(x)}{\Phi_0}\longrightarrow $}}\\
{\hspace{-0.0cm} (a) \hspace{9.0cm} (b)}
 \caption{The  NFW density $\frac{\rho}{\rho_0}$ as a function of the potential $\frac{\Phi}{\Phi_0}$ (a). The same quantity in the case  of the VO profile (b).  $\Phi_0=4 \pi G_Na^2 \rho_0 $ }
 \label{NFW_rhophi}
  \end{center}
  \end{figure}

\begin{figure}[!ht]
 \begin{center}
  \subfloat
 {
\rotatebox{90}{\hspace{-0.0cm} {$v_{rot}\longrightarrow \sqrt{4\pi G_Na^2 \rho_0}$}}
\includegraphics[scale=0.6]{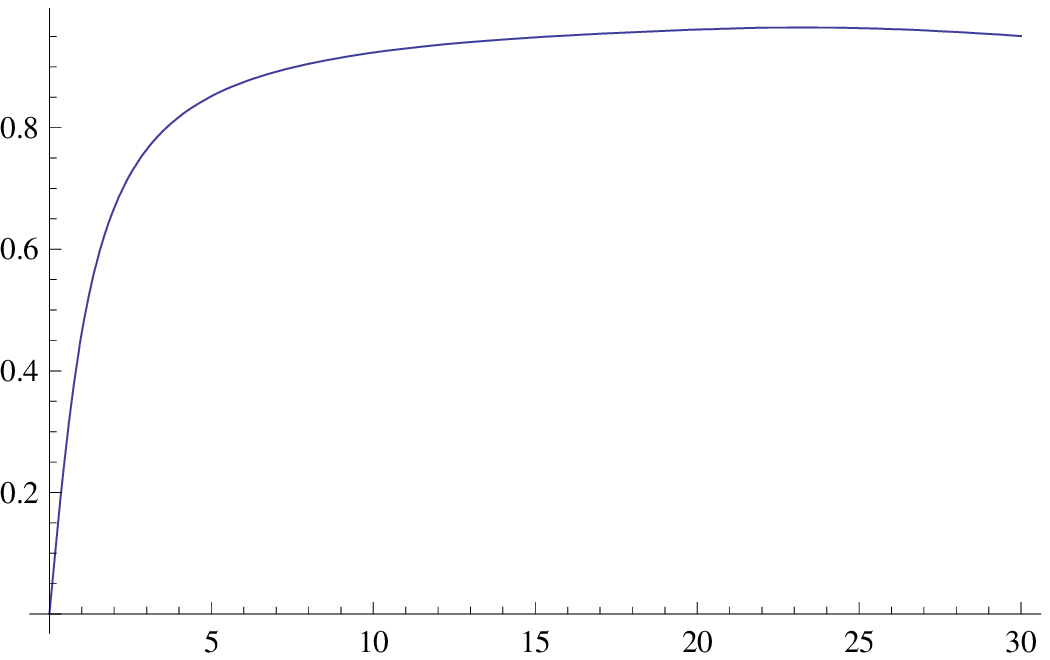}
}
 \subfloat
 { \rotatebox{90}{\hspace{-0.0cm} {$v_{rot}\longrightarrow
\sqrt{4\pi G_Na^2 \rho_0}$}}
\includegraphics[scale=0.6]{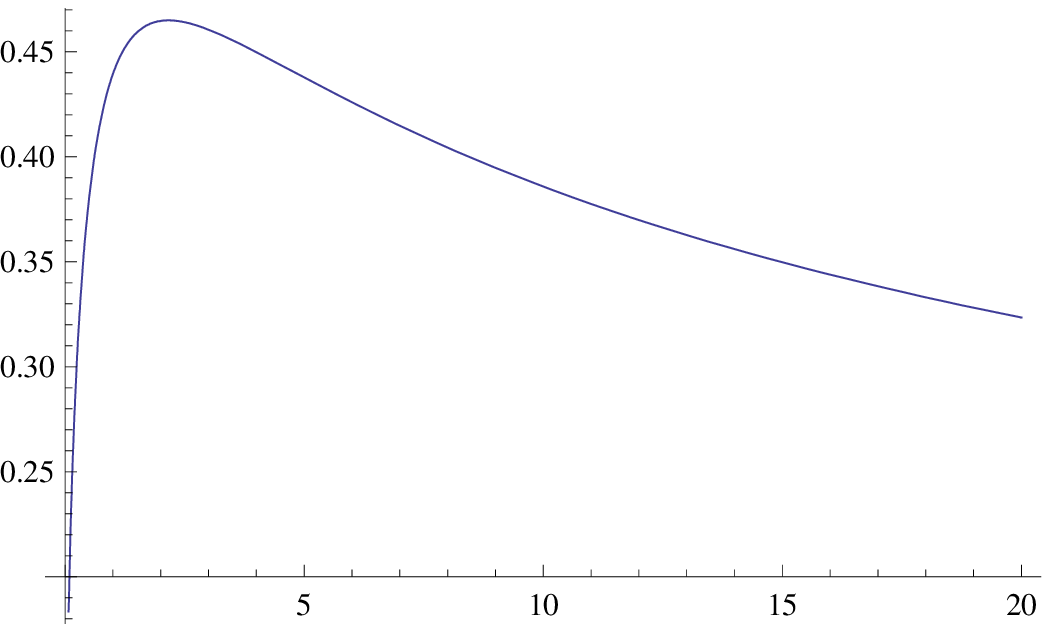}
}
{\hspace{-2.0cm}}\\
{\hspace{0.0cm} $r \longrightarrow a$}\\
{\hspace{-0.0cm} (a) \hspace{9.0cm} (b)}
 \caption{The rotational velocity due to dark matter as a function of the distance in units
 of $\sqrt{4\pi G_Na^2 \rho_0}$. Shown on the left is the one obtained with the
  VO density profile of Eq. (\ref{eq:VOprofile}), while on the right the
  NFW profile \cite{ULLIO},
  see Eq. (\ref{dens5b}), was employed.}
 \label{rotvel}
  \end{center}
  \end{figure}

There are many halo density profiles, which have been employed , see e.g. a recent summary \cite{KazZenKra06} and references therein.  Among the most commonly used analytic profiles \cite{ZHAO96} we will consider the following:
\begin{itemize}
\item The VO density profile \cite{VEROW06}
\beq
\rho(x)=\rho_0\left (\begin{array}{c}
\frac{1}{1+x^2,~},~ x\le c\\
 \left(\frac{2
   \left(c^2+1\right)}{\left(x^2+1\right)^2}-\frac{\left(c^2
   +1\right)^2}{\left(x^2+1\right)^3}\right),~x>c 
\end{array}
\right)
\label{eq:VOprofile}
\eeq
with $a$ the radius of the Galaxy. The distance $c$ is very large, so that the  the rotational velocity remains essentially constant with the distance
from the center of the galaxy even at quite large distances. The above form was taken by the requirement that at $x=c$ the density is continuous with a continuous derivative. This will be referred to as VO profile. The resulting potential can be obtained by solving Poisson's equation \cite{VEROW06}.
 \item Another simple profile is:
\begin{equation}
\rho(x)=\frac{\rho_0}{x(1+x)^2},~~ x=\frac{r}{a}
 \label{dens5b}
\end{equation}
known as NFW distribution \cite{NFW},\cite{ULLIO}, which has been suggested by  N-body simulations . This profile
provides a better description of the expected density near the
center of the galaxy. It does not, however, predict the constancy
of the rotational velocities at large distances.
\end{itemize}
As we have seen above the dependence of the density on the potential  is much more interesting. With thus show these functions in Figs 
\ref{NFW_rhophi}.
The obtained rotational velocity curve, due to dark mater alone, is given in Fig. \ref{rotvel} in units of $\sqrt{\Phi_0}$ with
$\Phi_0=4\pi G_Na^2 \rho_0$. One can obtain the value of $\rho_0$ by fitting the rotational velocity in our vicinity to the sun's rotational velocity 220 km~s$^{-1}$. Using $a=3.1\times 10^{20}$ m and $x_s=0.8$, i.e. our location $r_s=0.8a$, we find 0.3 and 0.5 GeV/$c^2$cm$^{-3}$ for the VO and NFW profiles respectively compared to the canonical value \cite{PDG} of 0.3 GeV/$c^2$cm$^{-3}$.

\section{Velocity Distributions}
Before proceeding further we define the asymmetry in the usual way:
\beq
\beta=1-\frac{\upsilon_t^2}{2 \upsilon^2_r}
\eeq
and it is usually assumed to be positive, $0\le \beta\le 0.5$. The asymmetry in our vicinity $r=r_s$  is related to the parameter $r_0$ introduced above via the relation. 
\beq
r_0=r_s\sqrt{\frac{1-\beta}{\beta}}
\eeq
 Substitute  the last expression in $\tilde{\rho}(\Phi,r_0)$ we get $\tilde{\rho}(\Phi,\beta)$. Viewed as a function of the potential $\tilde{\rho}(\Phi,\beta)$ and its derivatives up to third order vanish at the zero of the potential.

\begin{figure}[!ht]
 \begin{center}
  \subfloat
 {
\rotatebox{90}{\hspace{-0.0cm} {$f_{\upsilon} \longrightarrow$}}
\includegraphics[scale=0.6]{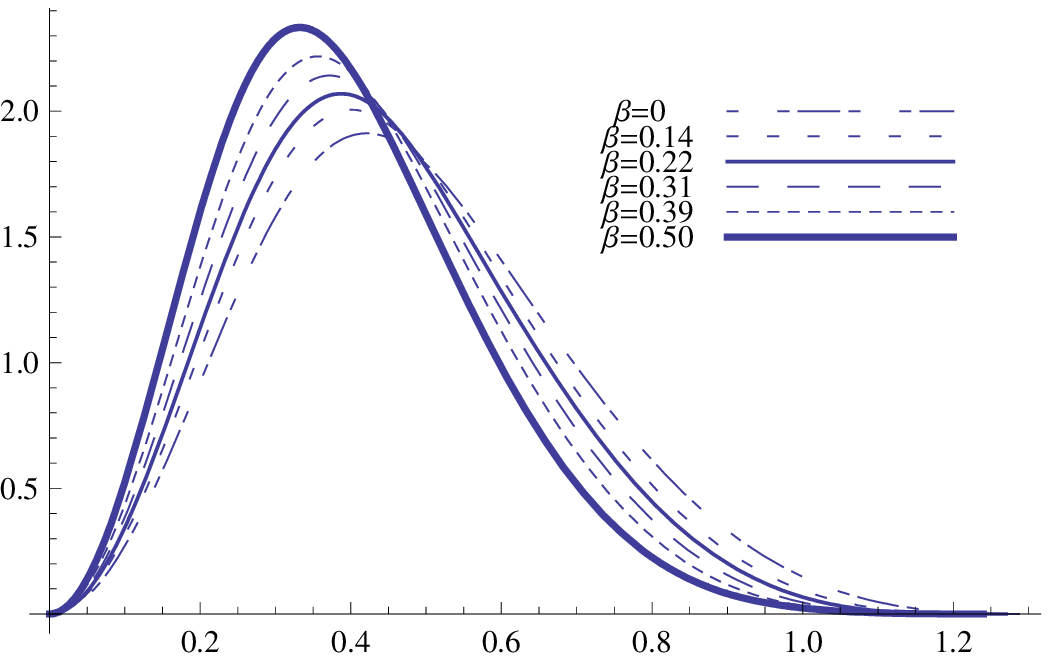}
}
  \subfloat
 {
\rotatebox{90}{\hspace{-0.0cm} {$f_{\upsilon} \longrightarrow$}}
\includegraphics[scale=0.6]{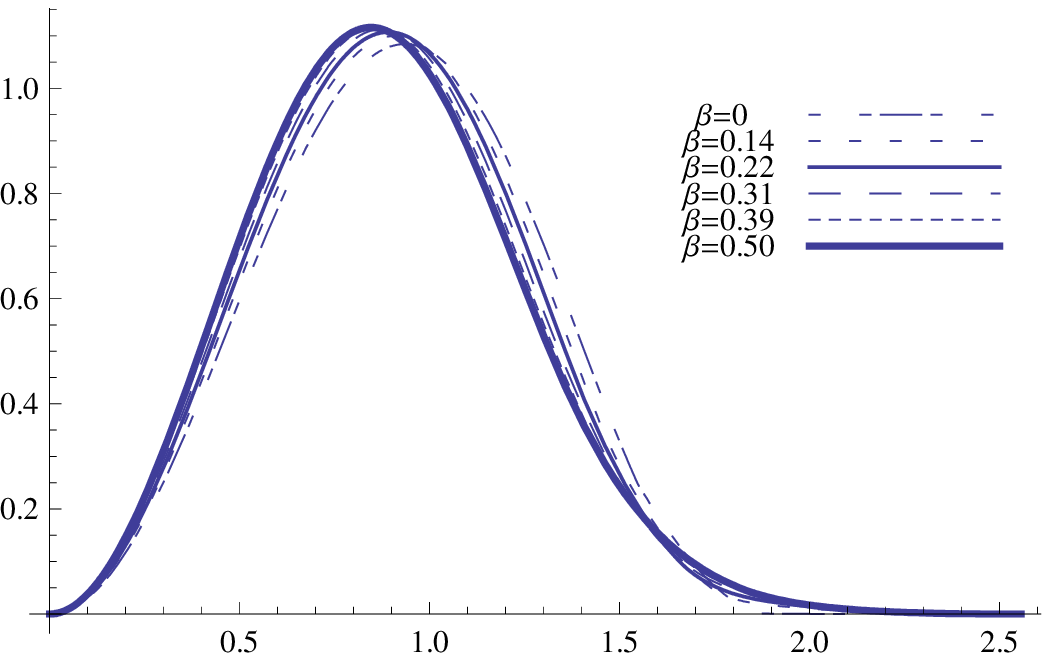}
}
{\hspace{-2.0cm}}\\
{\hspace{-2.0cm} $\frac{\upsilon}{\sqrt{|\Phi_0|}} \longrightarrow$}\\
{\hspace{-0.0cm} (a) \hspace{9.0cm} (b)}
 \caption{The  velocity distribution in our vicinity, $4 \pi y^2 f_{\upsilon}(y,\beta) $ obtained with the NFW profile as a function of the velocity in units of $ \sqrt{|\Phi_0|}=5.2\times 10^{3}$ km s $^{-1}$ on the left and the VO profile with $\sqrt{\Phi_0}=2.67\times 10^{3}$ km s $^{-1}$ on the right. In the plots the thick, dot-dot,long dash,fine continuous, short dash and dot-dash  lines correspond to $\beta=0.50, 0.39, 0.31, 0.22$  $0.14$  and $0$ respectively.
 }
 \label{fig:NFWfv_as}
  \end{center}
  \end{figure}

\begin{figure}[!ht]
 \begin{center}
  \subfloat
 {
\rotatebox{90}{\hspace{-0.0cm} {$f_{\upsilon} \longrightarrow$}}
\includegraphics[scale=0.6]{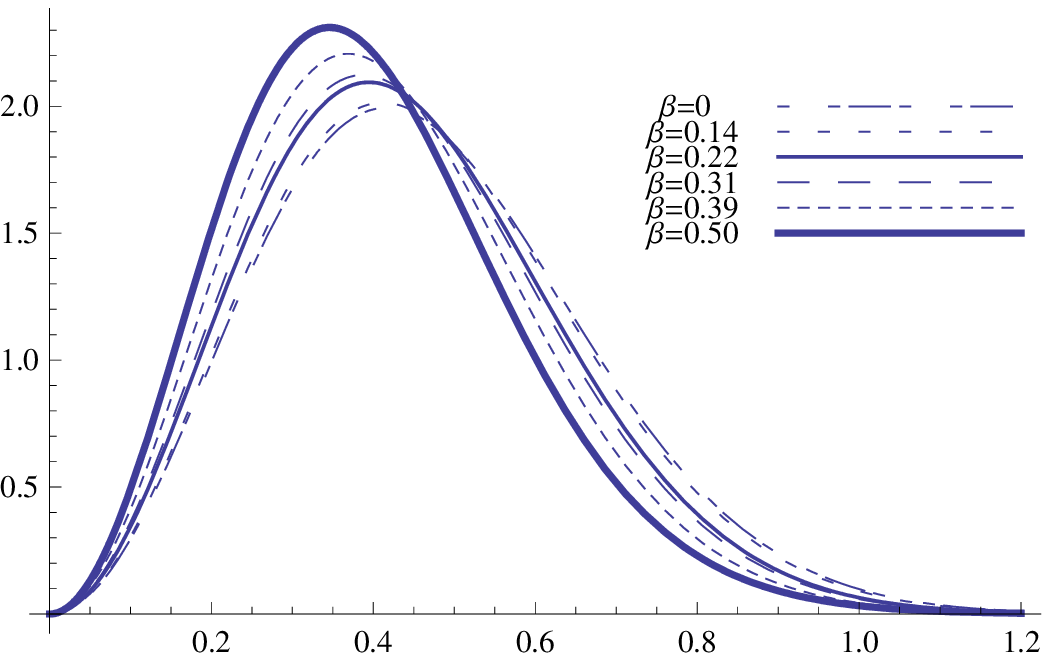}
}
  \subfloat
 {
\rotatebox{90}{\hspace{-0.0cm} {$f_{\upsilon} \longrightarrow$}}
\includegraphics[scale=0.6]{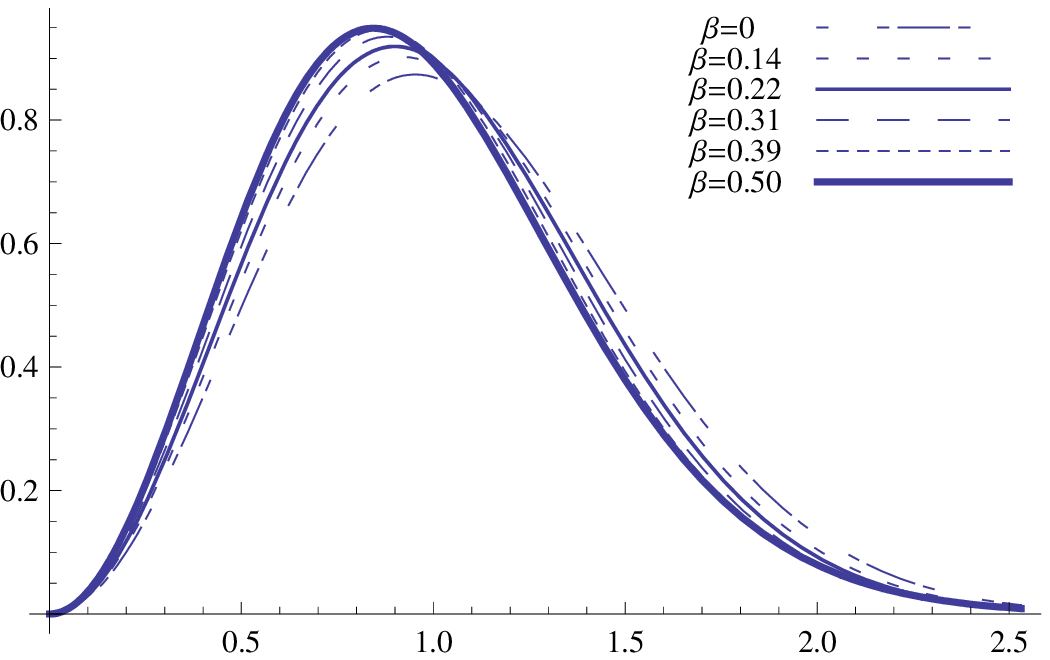}
}
{\hspace{-2.0cm}}\\
{\hspace{-2.0cm} $\frac{\upsilon}{\sqrt{|\Phi_0|}} \longrightarrow$}\\
{\hspace{-0.0cm} (a) \hspace{9.0cm} (b)}
 \caption{The same as in Fig. \ref{fig:NFWfv_as} but using the  approximate M-B  velocity distribution.}
 \label{NFWfv_mbas}.
  \end{center}
  \end{figure}

The velocity distribution can be cast in the form:
\beq
f_{r_s}(y,\xi,\beta)=f\left (\Phi(r_s)+\frac{1}{2}y^2  \frac{1-\beta \xi^2}{1-\beta} \right ),~~ y=\frac{\upsilon}{\sqrt{|\Phi_0|}},~~\xi=\cos{\theta}
\label{Eq:asveldis}
\eeq
The angle $\theta$ here is defined with respect to a polar axis taken in the direction of the radial component of the velocity.
\\ To get an idea of the effect of the asymmetry we integrate the normalized function over the angles. We thus get the  distributions shown in Fig. \ref{fig:NFWfv_as}. We see that the maximum allowed velocity for the NFW distribution we find $\sqrt{\Phi_0}=5.2\times 10^3$ km/s, i.e. is $\upsilon_{max}=1.21 Sqrt{|\Phi_0|}=1.21 5.2 \times 10^3$ km/s=$6.3\times 10^3$km/s, which leads to $\upsilon_{esc}=2.86\upsilon_0$, almost the same with the value previously used. On the other hand in the case of the VO profile  one finds \cite{VEROW06} $\sqrt{|\Phi_0|}=2.67\times 10^3$ km/s, i.e.  $\upsilon_{esc}=3.4 \upsilon_0$.
\\Since, as we have already mentioned, in many dark matter calculations of the event rates the simple symmetric or asymmetric  M-B velocity distribution has been employed, one would like to know to what extend this can be consistent with standard density profiles \cite{KazMagMoore04}. In the present work we would like, in particular, to know to what extent such an approximate axially symmetric M-B distribution, which vanishes outside the range of $y$ given by the exact distribution, can be derived in the context of the above density profiles, and can then be  used in calculations related to dark matter searches. We seek a solution of the form: 
\beq
f_v^{MB}=C e^{-\frac{y^2 \left(1-\frac{2 \beta }{3}\right)
   \left(1-\beta  \xi ^2\right)}{b^2 (1-\beta )}}
\label{Eq:MBdis}
\eeq
where $C$ is  a normalization constant. The above form has been motivated not only by the structure of Eq. (\ref{Eq:asveldis}), but in addition by our previous work, in which this shape  has been obtained by an appropriate limit of velocity distributions described by the radial and tangential Tsallis type functions \cite{VerHanH}. The width of the M-B distribution 
depends on the assumed density profile, following the trend of the exact distributions. The parameter $b$ and the allowed range of $y$, which are functions of $\beta$ are shown in table \ref{tab:MBdata}.
\begin{table}[ht!]
\begin{center}
\caption{ We show the parameters describing the M-B distribution, which is a good fit to the velocity distribution derived from the NFW and VO dark matter density profiles via the Eddington approach. The range of y given  is in units of $\sqrt|\Phi_0|$.
\label{tab:MBdata}}
\begin{tabular}{|c||c|c|c|c|c|c|c|c|}
\hline
\hline 
&$\beta$&0.5& 0.39& 0.31& 0.22& 0.14&0.00\\
\hline
NFW&a&0.356& 0.403& 0.437& 0.471& 0.503& 0.555\\
\hline
NFW&$y_{max}$&1.21&1.21&1.21&1.21&1.21&1.22\\
\hline
\hline
VO&a&0.758&0.815&0.862&0.912&0.962&1.044\\
\hline
VO&$y_{max}$&2.70&2.70&2.70&2.70&2.70&2.70\\
\hline
\hline
\end{tabular}
\end{center}
\end{table}
\\ Integrating the distribution of Eq. (\ref{Eq:MBdis}) over the angles we obtain the results  shown in Fig. \ref{NFWfv_mbas}.
\section{transformation into our local coordinates.}
We must now transform the above distributions from the galactic to the local coordinates.
\beq
f_{r_s}(y,\xi,\beta)\rightarrow f^{\mbox{local}}_{\upsilon}(y,\theta,\phi,\beta,\alpha,\delta,\gamma)
\eeq
 This is accomplished by the substitutions:
\beq
y^2\rightarrow X^2+Y^2+Z^2,~~\xi^2\rightarrow \frac{X^2}{X^2+Y^2+Z^2}
\eeq
where
\barr
X&=&\frac{1}{sc}(y \cos \phi  \sin \theta +\delta  \sin \alpha  ),~~
Y=\frac{1}{sc}( y \sin \theta  \sin \phi -\delta  \cos \alpha  \cos \gamma ),
\nonumber\\
Z&=&\frac{1}{sc}(y \cos \theta +\delta  \cos \alpha  \sin \gamma +1)~~,~~y=\frac{\upsilon}{\upsilon_0}~~,~~
sc=\frac{\sqrt{2 |\Phi(x_s)|}}{\upsilon_{\mbox{rot}}(x_s)}\approx 2.86
\earr
The angles $\theta$ and $\phi$ are the spherical coordinates  defined in the usual way. Here the polar axis has been chosen along the sun's direction of motion, the x-axes radially out of the galaxy and the y-axis perpendicular in the galactic plane ($\hat{y}=\hat{z}\times \hat{x} $). $\delta=0.135$ is
the earth's rotational velocity in units of the sun's velocity, $\gamma \approx \pi$/6 is the angle between the axis of the galaxy and the axis of the ecliptic and $\alpha$ is the phase of the earth ($\alpha=0$ around June the 3nd).
 \section{A brief discussion of direct WIMP event rates (DWER)}
\label{sec:rates}
Even though the expressions for the event rates for WIMP detection are 
 well known for the reader's convenience we will include the basic formulas here in our own notation
 \cite{Verg01,JDV03,JDVSPIN04,JDV06a}:
\begin{equation}
 \Big<\frac{dR}{du}\Big> =
\frac{\rho (0)}{m_{\chi}} \frac{m}{Am_N} \sqrt{\langle
\upsilon^2\rangle }\int
           \frac{   |{\boldmath \upsilon}|}
{\sqrt{ \langle \upsilon^2 \rangle}}
f^{\mbox{local}}_{\upsilon}(\frac{\upsilon}{\upsilon_0},\theta,\phi,\beta,\alpha,\delta,\gamma)
                       \frac{d\sigma (u,\upsilon )}{du} d^3
 \mbox{\boldmath $\upsilon$}
\label{3.11}
\end{equation}

The differential cross section is given by \cite{JDV06a}:
\beq
d\sigma (u,\upsilon)= \frac{du}{2 (\mu _r b\upsilon )^2}
 [(\bar{\Sigma} _{S}F(u)^2
                       +\bar{\Sigma} _{spin} F_{11}(u)] 
\label{2.9}
\end{equation}
where $ u$ the energy transfer $Q$ in dimensionless units given by
\begin{equation}
 u=\frac{Q}{Q_0}~~,~~Q_{0}=[m_pAb]^{-2}=40A^{-4/3}~MeV
\label{defineu}
\end{equation}
 with  $b$ is the nuclear (harmonic oscillator) size parameter. $F(u)$ is the
nuclear form factor and $F_{11}(u)$ is the spin response function associated with
the isovector channel.

The scalar and spin cross sections are given by:
\begin{equation}
\bar{\Sigma} _S 
\approx  \sigma^{S}_{N,\chi^0} (\frac{\mu_r(A)}{\mu_r (p)})^2 A^2~~,~~\bar{\Sigma} _{spin}  =  (\frac{\mu_r(A)}{\mu_r(p)})^2
                           \sigma^{spin}_{N,\chi^0}~\zeta_{spin}
\label{2.10}
\end{equation}
where $A$ is the nuclear mass number and $\mu_r(A) $ ($\mu_r(p) $ ) is the WIMP-Nucleus (WIMP-nucleon) reduced mass.  $\sigma^S_{N,\chi^0}$ and  $ \sigma^{spin}_{N,\chi^0}$ are  respectively the WIMP-nucleon scalar and spin  cross sections and $~\zeta_{spin}$ the nuclear spin ME.

Integrating over the energy transfer $u$ we obtain the  event rate for  WIMP-nucleus elastic scattering, which is given by \cite{Verg01,JDV03,JDVSPIN04,JDV06a}:
\barr
R&=& \frac{\rho (0)}{m_{\chi^0}} \frac{m}{m_p}~
              \sqrt{\langle v^2 \rangle } 
\nonumber\\
& &\left [f_{coh}(A,\mu_r(A)) \sigma_{p,\chi^0}^{S}+f_{spin}(A,\mu_r(A))\sigma _{p,\chi^0}^{spin}~\zeta_{spin} \right]
\label{fullrate}
\earr
with
\beq
f_{coh}(A, \mu_r(A))=\frac{100\mbox{GeV}}{m_{\chi^0}}\left[ \frac{\mu_r(A)}{\mu_r(p)} \right]^2 A~t_{\mbox{coh}}\left(1+h_{\mbox{coh}}cos\alpha \right)
\label{Eq:thcoh}
\eeq
\beq
f_{spin}(A, \mu_r(A))=\left[ \frac{\mu_r(A)}{\mu_r(p)} \right]^2 \frac{t_{\mbox{spin}}\left(1+h_{\mbox{spin}}cos\alpha \right)}{A}
\eeq
The nucleon coherent and spin cross sections are the most important
particle physics parameters and $\zeta_{spin}$ is the most important nuclear physics parameter. In this work, however, we are mainly interested  in the effects of the velocity distribution on the event rate. So to spare the reader the inconvenience of detailed discussions of particle and nuclear physics we will not be concerned with the spin cross section. Thus we essentially only need the information of the velocity distribution and the effect of the nuclear form factor. Thus the
relevant parameters and $t_{coh}$, which deal with the time averaged rate, and $h_{coh}$, which  deals with the  modulation due to the annual  motion of the Earth. They result after the folding of the nuclear form factor with the WIMP
 velocity distribution. More specifically:
\beq
t_{coh}= \int_{u_{{min}}}^{u_{{max}}} \frac{dt_{coh}}{du} du~,~ h_{coh}=\frac{1}{t_{coh}} \int_{u_{{min}}}^{u_{{max}}}  \frac{dh_{coh}}{du} du.
\eeq
The energy transfer $u$ is limited
from below by the detector energy threshold and from above by the maximum WIMP velocity, i.e. :
\beq
u_{{min}}=\frac{Q_{{th}}}{Q_0}\leq u \leq u_{{max}}=\left( \frac{y_{{max}}}{a} \right)^2~,~a= \left[\sqrt{2}\mu_r b \upsilon_0 \right ]^{-1},~~\upsilon_0=\mbox{ the sun's velocity}
\eeq
One can show that
\begin{equation}
\frac{dt}{du}= \sqrt{\frac{2}{3}} a^2 F^2(u) \Psi_0(a \sqrt{u}),~~\frac{dh}{du}=\sqrt{\frac{2}{3}} a^2 F^2(u)H(a \sqrt{u}) \cos{\alpha}.
\label{eq:rrate}
\end{equation}
where the two functions $ \Psi_0(a \sqrt{u})$ and $H(a \sqrt{u}) $ are respectively the $n=0$ ($\alpha$ independent) and the  $\cos{\alpha}$ Fourier coefficients of the integral:
\beq
J=\int_{a \sqrt{u}}^{y_{max}} y dy \int f^{\mbox{local}}_{\upsilon}(y,\theta,\phi,\beta,\alpha,\delta,\gamma)  d \hat{y} 
\eeq
The integral $J$  contains all the relevant information  on the velocity distribution. We see from this that essentially the first moment of the velocity distribution appears. Since the high energy transfers are suppressed by the nuclear form factor $F(u)$, the above integral is not much affected by the behavior of the distribution near the escape velocity. So we expect the M-B approximation to be an adequate description of the exact distribution.

The above expressions manifestly show the essential parts:
\begin{itemize}
\item The elementary cross section.\\
This is the most important part. It depends on two ingredients i) The particle model which provides the amplitude at the quark level and ii) The procedure for going from the quark to the nucleon level. We will not concern ourselves with such issues here.
\item The WIMP density in our vicinity $\rho_0$.\\
This has been obtained in two phenomenological density profile as described above.
\item The  dependence of the rate on the cross properties of the target and the WIMP mass. 
\item The quantity $t$.\\
This is independent of the parameters of the particle model, except for the WIMP mass. It takes into account:
\begin{enumerate}
\item The nuclear structure effects (for the coherent process the nuclear form factor).
\item The WIMP velocity distribution.\\
This may be obtained from a given spherical density  profile via the Eddington approach as discussed above.
\item The energy threshold imposed by the detector. In the case of a non zero threshold the obtained rates depend on quenching. Such an effect will not be discussed here. The interested reader is referred to the literature (see e.g. \cite{VEREJ08}-\cite{FS08a})
\end{enumerate}
\item The quantity $h$.\\
This describes the modulation of the amplitude due to  the  Earth's motion around the sun. It depends on the same parameters as $t$. 
\end{itemize}
 The evaluation of the quantities $t$ and $h$ with the obtained asymmetric velocity distribution will be discussed below.

 With the above ingredients the number of events in time $t$ due to the scalar interaction, which leads to coherence  \cite{TETRVER06}, can be cast in the form
\barr
 R&&\simeq  1.60~10^{-3}\times 
 \nonumber\\
 &&
\frac{t}{1 \mbox{y}} \frac{\rho(0)}{ \mbox {0.3GeVcm}^{-3}}
\frac{m}{\mbox{1kg}}\frac{ \sqrt{\langle
v^2 \rangle }}{280 {\mbox {km s}}^{-1}}\frac{\sigma_{p,\chi^0}^{S}}{10^{-6} \mbox{ pb}} f_{coh}(A, \mu_r(A))
\label{eventrate}
\earr
where the elementary cross section $ \sigma_{p,\chi^0}^{S}$can be treated as a phenomenological parameter.
\section{Results}
\begin{figure}[!ht]
 \begin{center}
  \subfloat
 {
\rotatebox{90}{\hspace{-0.0cm} {$\frac{dt_{coh}}{du}
\longrightarrow$}}
\includegraphics[scale=0.6]{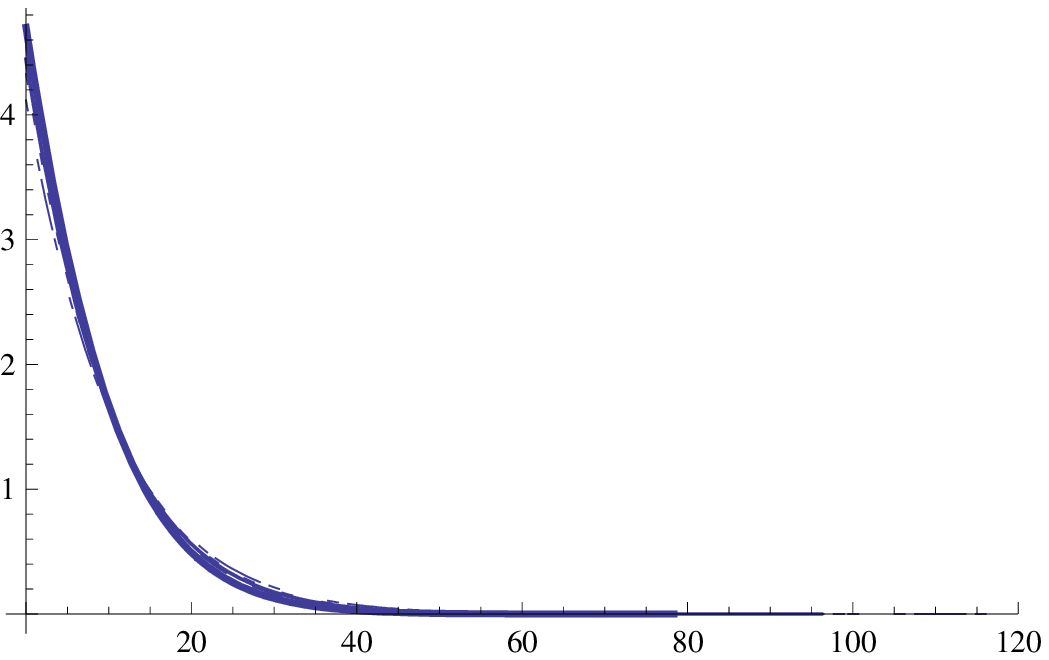}
}
 \subfloat
 {
\rotatebox{90}{\hspace{-0.0cm} {$\frac{dt_{coh}}{du}
\longrightarrow$}}
\includegraphics[scale=0.6]{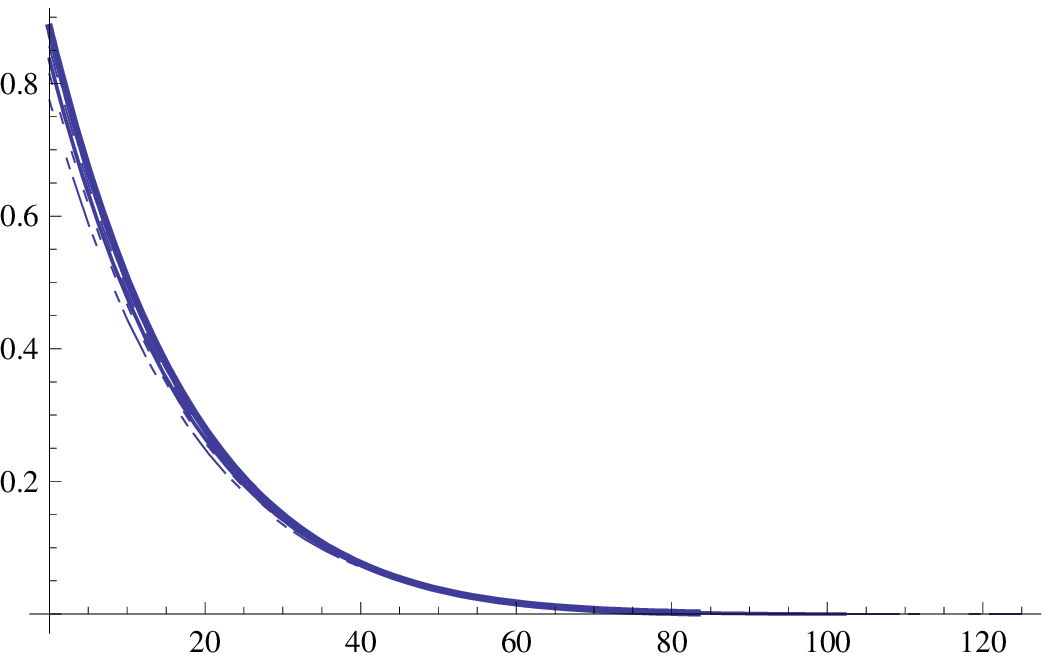}
}\\
{\hspace{-2.0cm}}
{\hspace{-2.0cm}}
{\hspace{5.0cm} $Q\longrightarrow$  keV}\\
{\hspace{-0.0cm} (a) \hspace{9.0cm} (b)}
 \caption{On the left we show the differential rate $ \frac{dt_{coh}}{du}$ for $m_{\chi}=30$ GeV
 in the case of the axially symmetric distribution obtained in this work for a heavy nucleus. Here we have taken  A=127. On the right we show
 the same quantity for $m_{\chi}=100$ GeV. Otherwise the notation is the same as in Fig. \ref{fig:NFWfv_as}. The effect of the asymmetry is not visible. For the spin contribution the behavior is analogous.} 
 \label{Fig:r127}
   \end{center}
  \end{figure}

\begin{figure}[!ht]
 \begin{center}
  \subfloat
 {
\rotatebox{90}{\hspace{-0.0cm} {$\frac{dh_{coh}}{du}
\longrightarrow$}}
\includegraphics[scale=0.6]{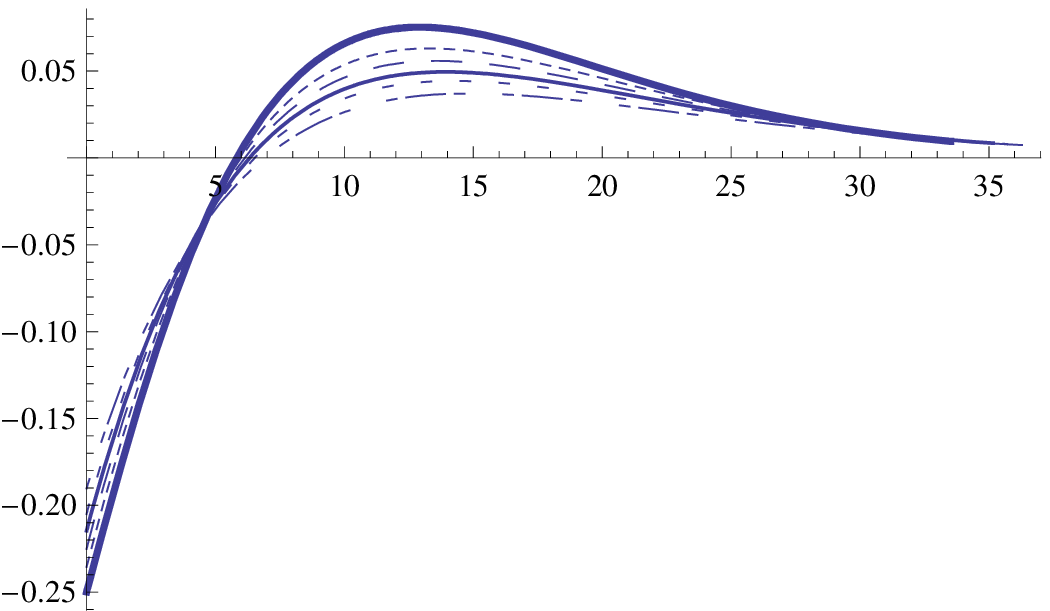}
}
 \subfloat
 {
\rotatebox{90}{\hspace{-0.0cm} {$\frac{dh_{coh}}{du}
\longrightarrow$}}
\includegraphics[scale=0.6]{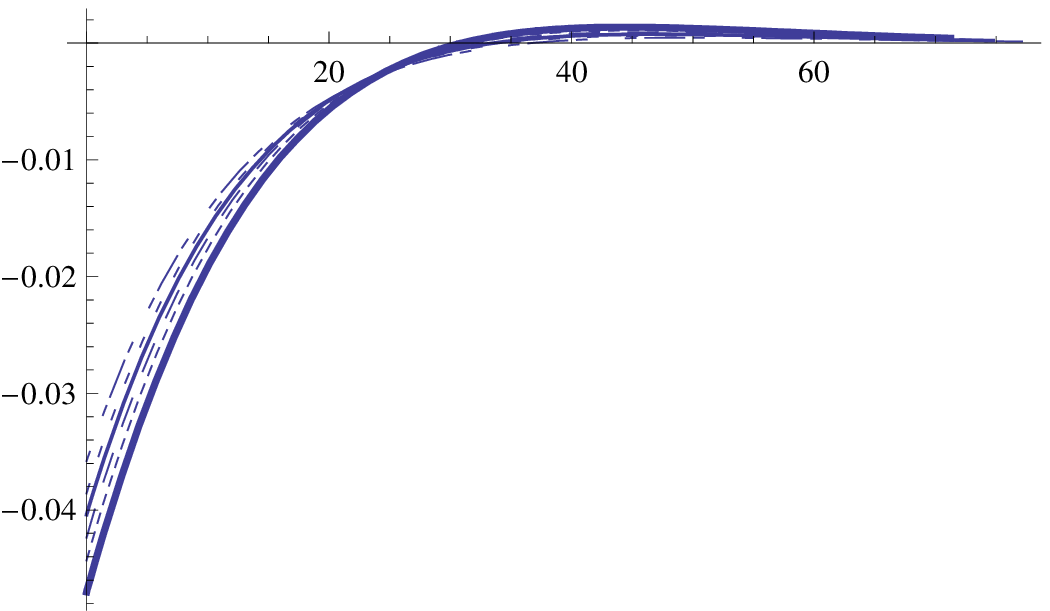}
}\\
{\hspace{-2.0cm}}
{\hspace{-2.0cm}}
{\hspace{5.0cm} $Q\longrightarrow$  keV}\\
{\hspace{-0.0cm} (a) \hspace{9.0cm} (b)}
 \caption{The same as in Fig. \ref{Fig:r127} for the quantity $ \frac{dh_{coh}}{dQ}$.}
 \label{Fig:h127}
   \end{center}
  \end{figure}

\begin{figure}[!ht]
 \begin{center}
  \subfloat
 {
\rotatebox{90}{\hspace{-0.0cm} {$\frac{dt_{coh}}{du}
\longrightarrow$}}
\includegraphics[scale=0.6]{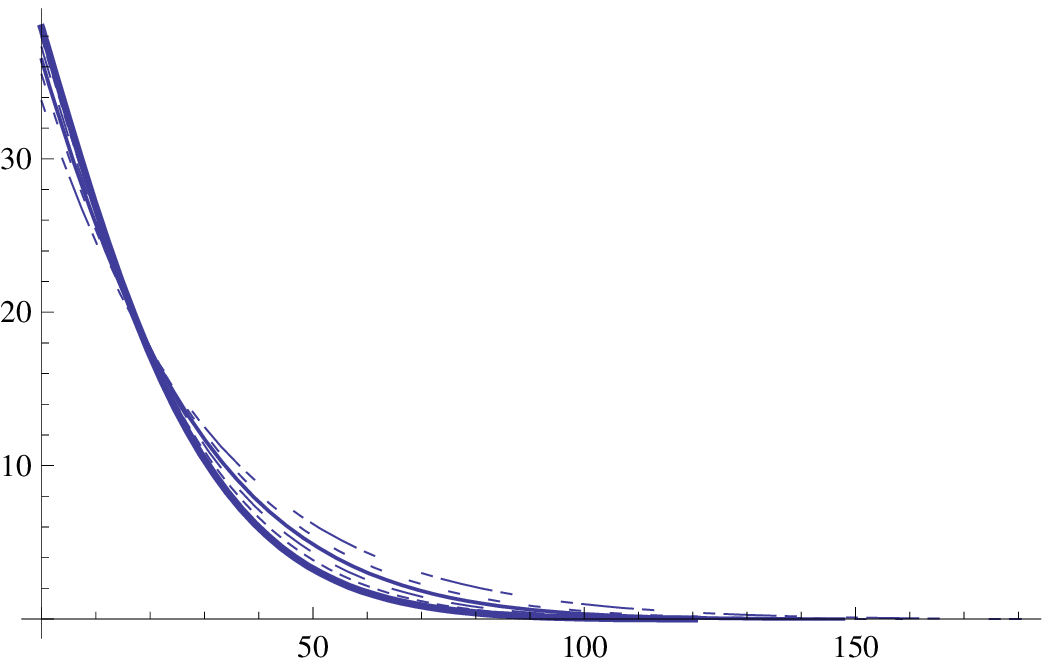}
}
 \subfloat
 {
\rotatebox{90}{\hspace{-0.0cm} {$\frac{dt_{coh}}{dQ}
\longrightarrow$}}
\includegraphics[scale=0.6]{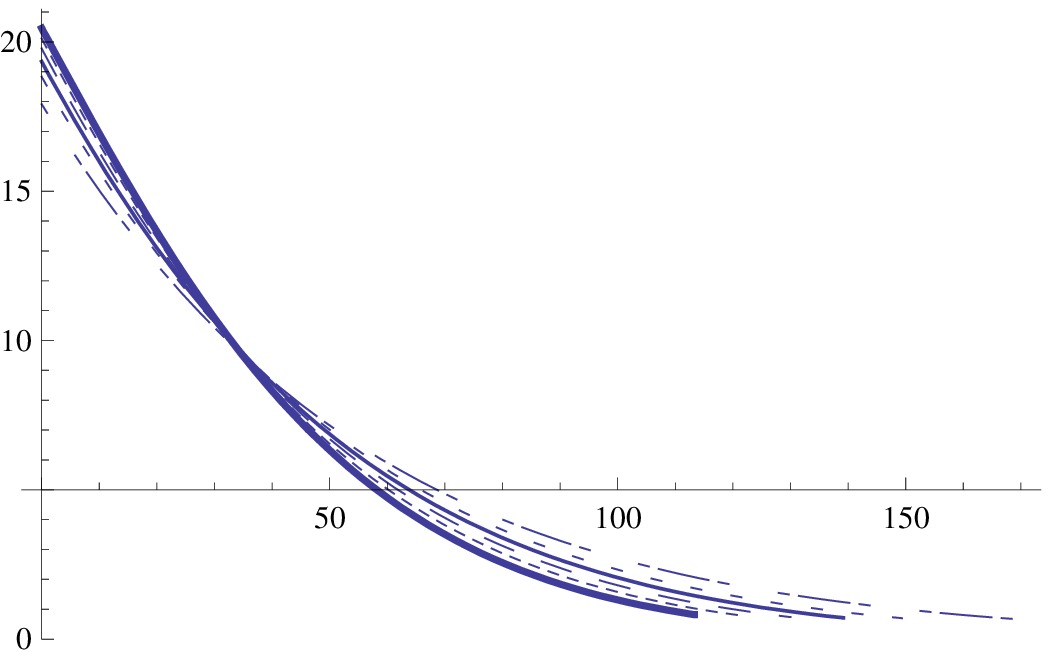}
}\\
{\hspace{-2.0cm}}
{\hspace{-2.0cm}}
{\hspace{5.0cm} $Q\longrightarrow$  keV}\\
{\hspace{-0.0cm} (a) \hspace{9.0cm} (b)}
 \caption{On the left we show the quantity $ \frac{dt_{coh}}{du}$ for $m_{\chi}=30$ GeV
 in the case of the axially symmetric distribution obtained in this work for a light nucleus like A=19. On the right we show
 the same quantity for $m_{\chi}=100$ GeV. Otherwise the notation is the same as in Fig. \ref{fig:NFWfv_as}.  For the spin contribution the behavior is analogous.} 
 \label{Fig:r19}
   \end{center}
  \end{figure}

\begin{figure}[!ht]
 \begin{center}
  \subfloat
 {
\rotatebox{90}{\hspace{-0.0cm} {$\frac{dh_{coh}}{du}
\longrightarrow$}}
\includegraphics[scale=0.6]{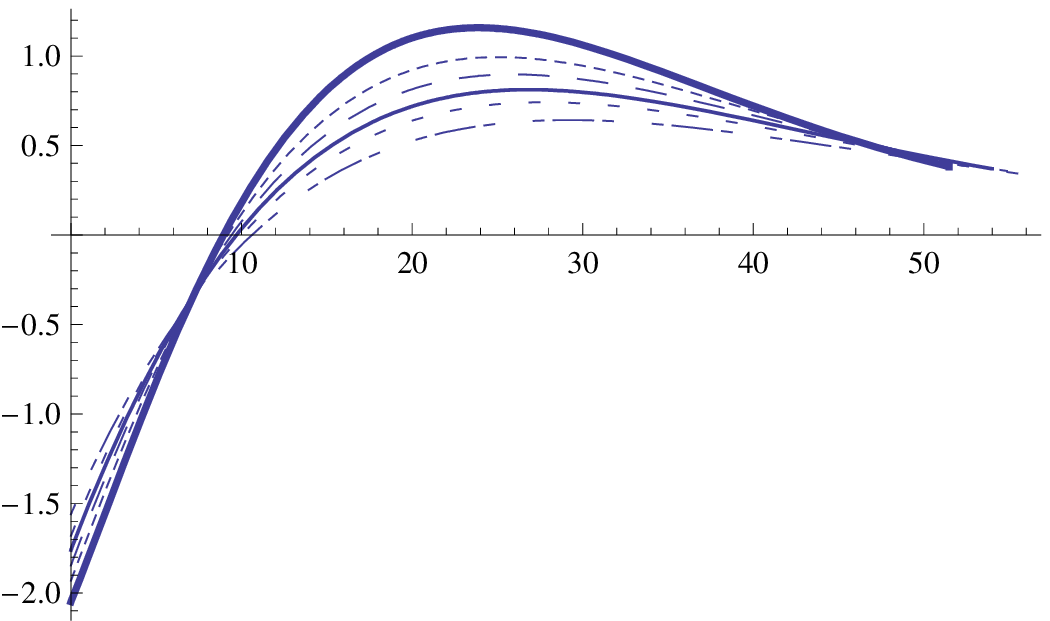}
}
 \subfloat
 {
\rotatebox{90}{\hspace{-0.0cm} {$\frac{dh_{coh}}{du}
\longrightarrow$}}
\includegraphics[scale=0.6]{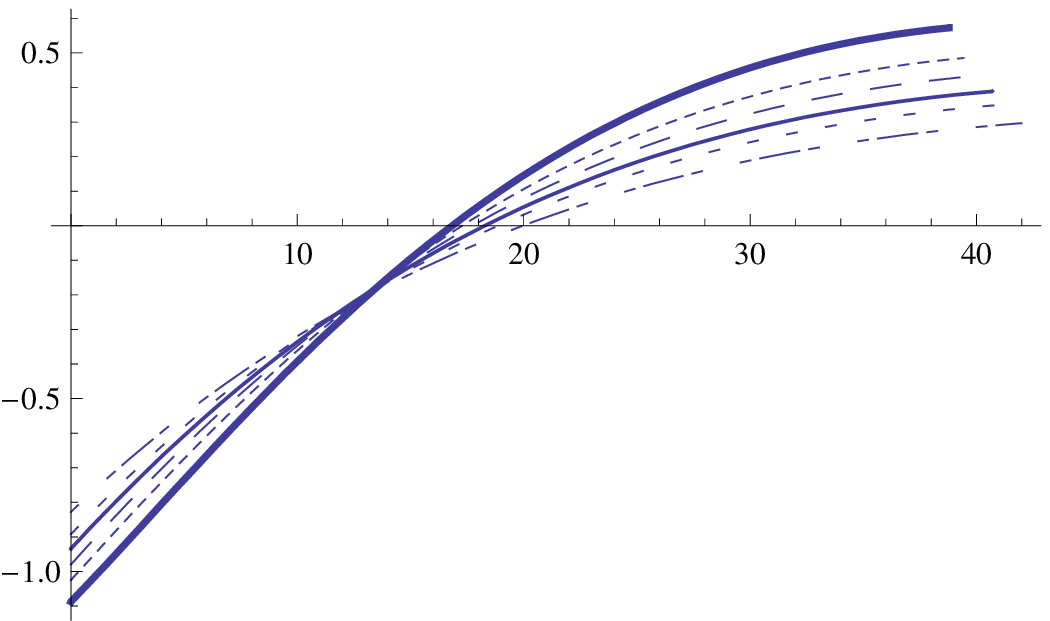}
}\\
{\hspace{-2.0cm}}
{\hspace{-2.0cm}}
{\hspace{5.0cm} $Q\longrightarrow$  keV}\\
{\hspace{-0.0cm} (a) \hspace{9.0cm} (b)}
 \caption{The same as in Fig. \ref{Fig:r19} for the quantity $ \frac{dh_{coh}}{du}$.}
 \label{Fig:h19}
   \end{center}
  \end{figure}

\begin{figure}[!ht]
 \begin{center}
  \subfloat
 {
\rotatebox{90}{\hspace{-0.0cm} {$t_{coh} 
\longrightarrow$}}
\includegraphics[scale=0.6]{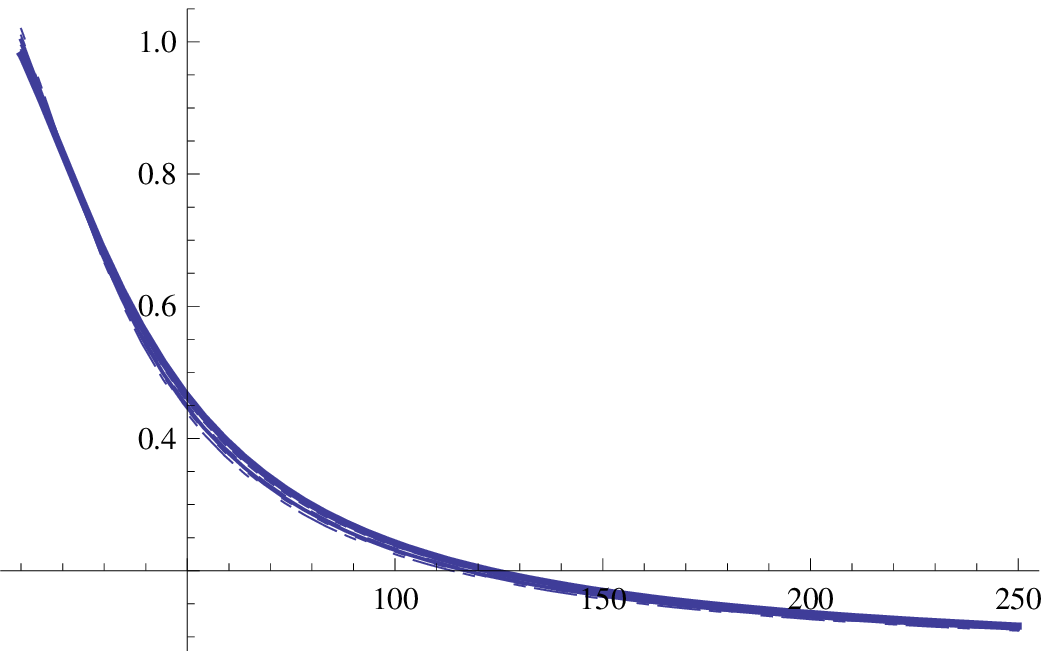}
}
 \subfloat
 {
\rotatebox{90}{\hspace{-0.0cm} {$t_{coh}
\longrightarrow$}}
\includegraphics[scale=0.6]{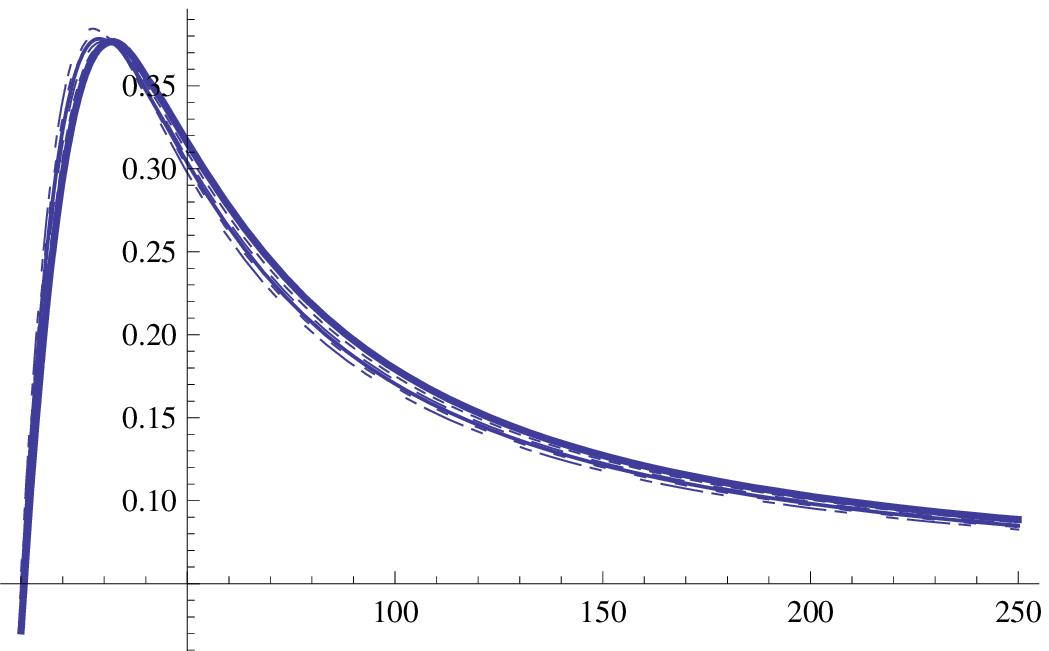}
}\\
{\hspace{-2.0cm}}
{\hspace{-2.0cm}}
{\hspace{5.0cm} $m_{\chi}\longrightarrow$  GeV}\\
{\hspace{-0.0cm} (a) \hspace{9.0cm} (b)}
 \caption{On the left we show the  rate $ {t_{coh}}$ as a function of the WIMP mass
 in the case of the axially symmetric distribution obtained in this work for a heavy nucleus. Here we have taken  A=127 and assumed zero threshold. On the right we show
 the same quantity for a threshold energy  of 5 keV, Otherwise the notation is the same as in Fig. \ref{fig:NFWfv_as}.}
 \label{Fig:tr127}
   \end{center}
  \end{figure}

\begin{figure}[!ht]
 \begin{center}
  \subfloat
 {
\rotatebox{90}{\hspace{-0.0cm} {$h_{coh}
\longrightarrow$}}
\includegraphics[scale=0.6]{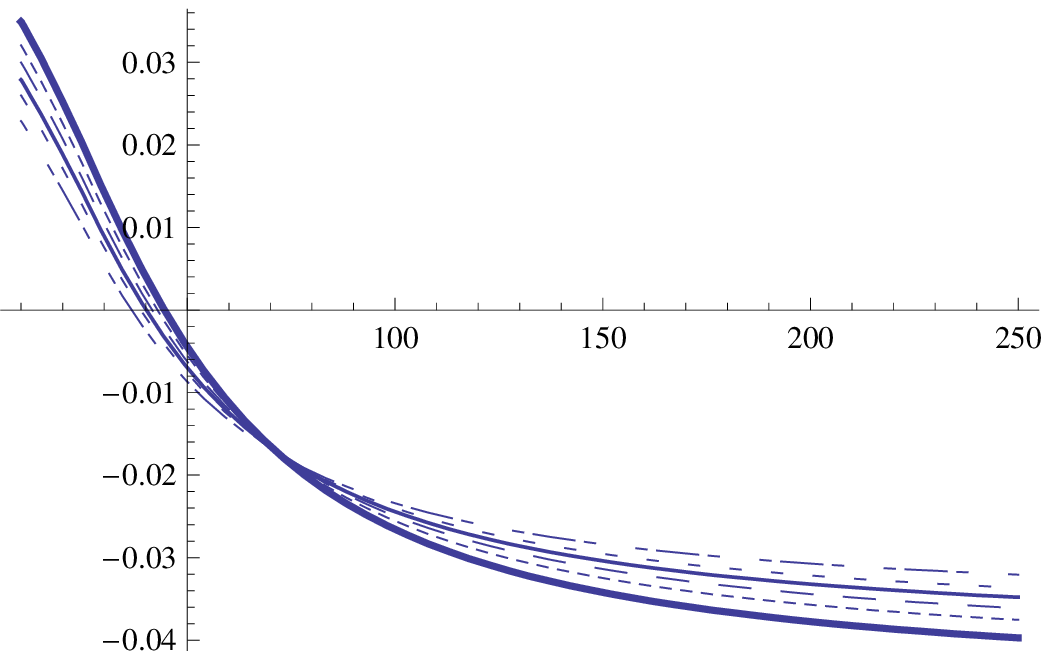}
}
 \subfloat
 {
\rotatebox{90}{\hspace{-0.0cm} {$h_{coh}
\longrightarrow$}}
\includegraphics[scale=0.6]{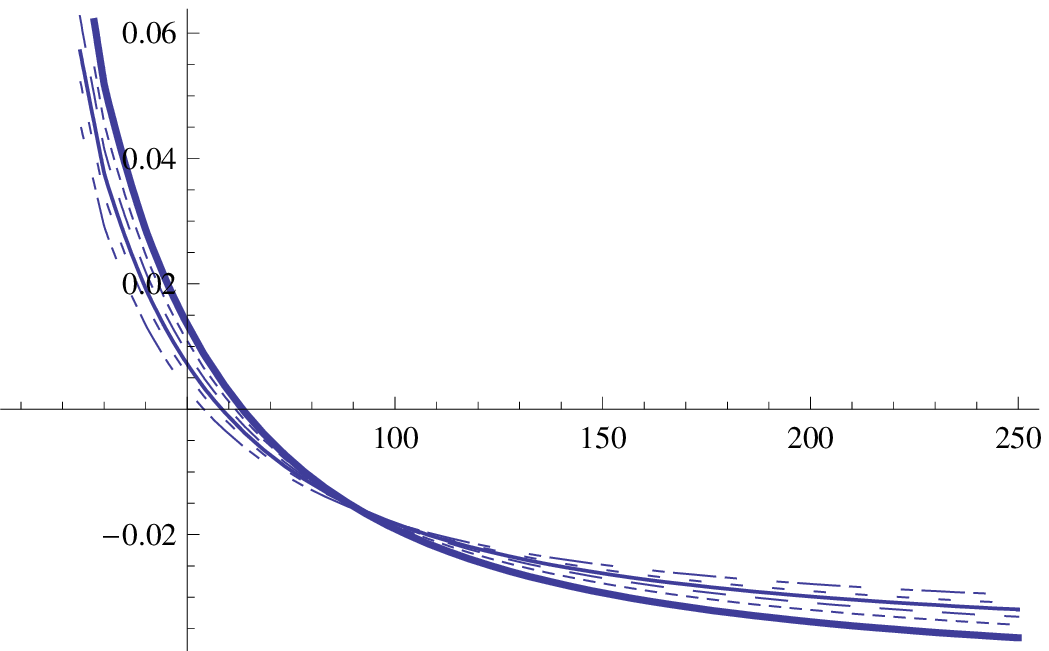}
}\\
{\hspace{-2.0cm}}
{\hspace{-2.0cm}}
{\hspace{5.0cm} $m_{\chi}\longrightarrow$  GeV}\\
{\hspace{-0.0cm} (a) \hspace{9.0cm} (b)}
 \caption{The same as in Fig. \ref{Fig:tr127} for the quantity $ {h_{coh}}$.}
 \label{Fig:th127}
   \end{center}
  \end{figure}

\begin{figure}[!ht]
 \begin{center}
  \subfloat
 {
\rotatebox{90}{\hspace{-0.0cm} {$t_{coh}
\longrightarrow$}}
\includegraphics[scale=0.6]{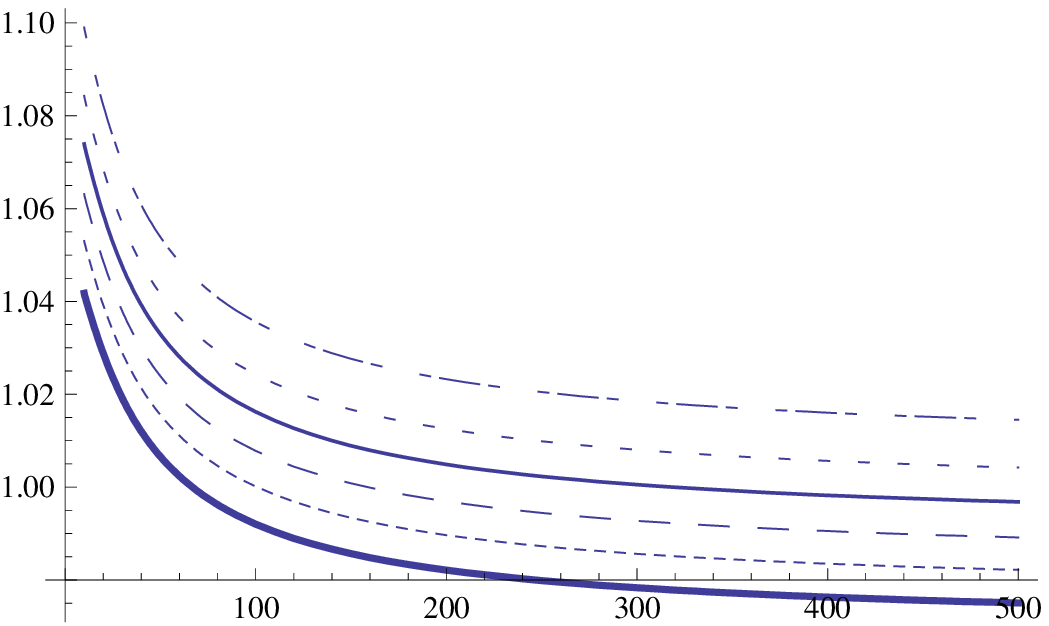}
}
 \subfloat
 {
\rotatebox{90}{\hspace{-0.0cm} {$t_{coh}
\longrightarrow$}}
\includegraphics[scale=0.5]{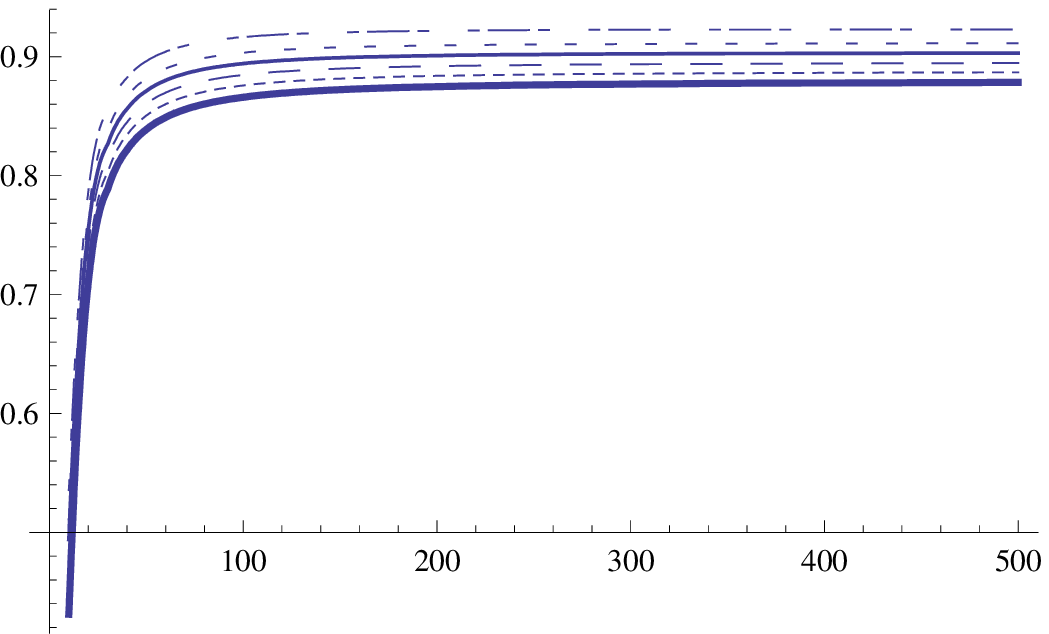}
}\\
{\hspace{-2.0cm}}
{\hspace{-2.0cm}}
{\hspace{5.0cm} $m_{\chi}\longrightarrow$  GeV}\\
{\hspace{-0.0cm} (a) \hspace{9.0cm} (b)}
 \caption{On the left we show the quantity $ {t_{coh}}$ as a function of the WIMP mass in  GeV
 in the case of the axially symmetric distribution obtained in this work for a light nucleus like A=19. On the right we show
 the same quantity for a threshold energy of 5  keV. Otherwise the notation is the same as in Fig. \ref{fig:NFWfv_as}.} 
 \label{Fig:tr19}
   \end{center}
  \end{figure}

\begin{figure}[!ht]
 \begin{center}
  \subfloat
 {
\rotatebox{90}{\hspace{-0.0cm} {$h_{coh}
\longrightarrow$}}
\includegraphics[scale=0.6]{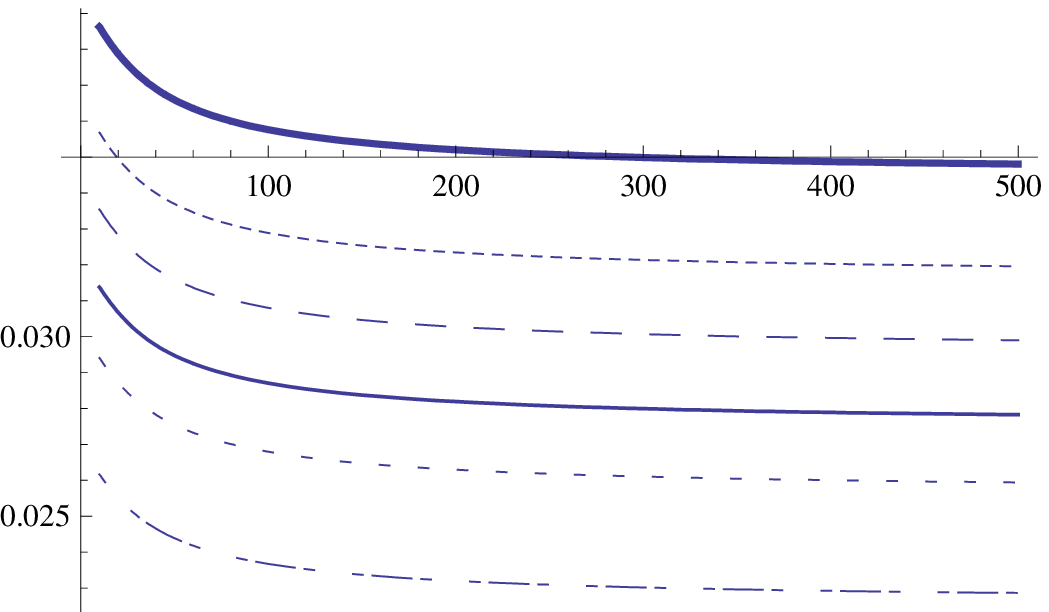}
}
 \subfloat
 {
\rotatebox{90}{\hspace{-0.0cm} {$h_{coh}
\longrightarrow$}}
\includegraphics[scale=0.6]{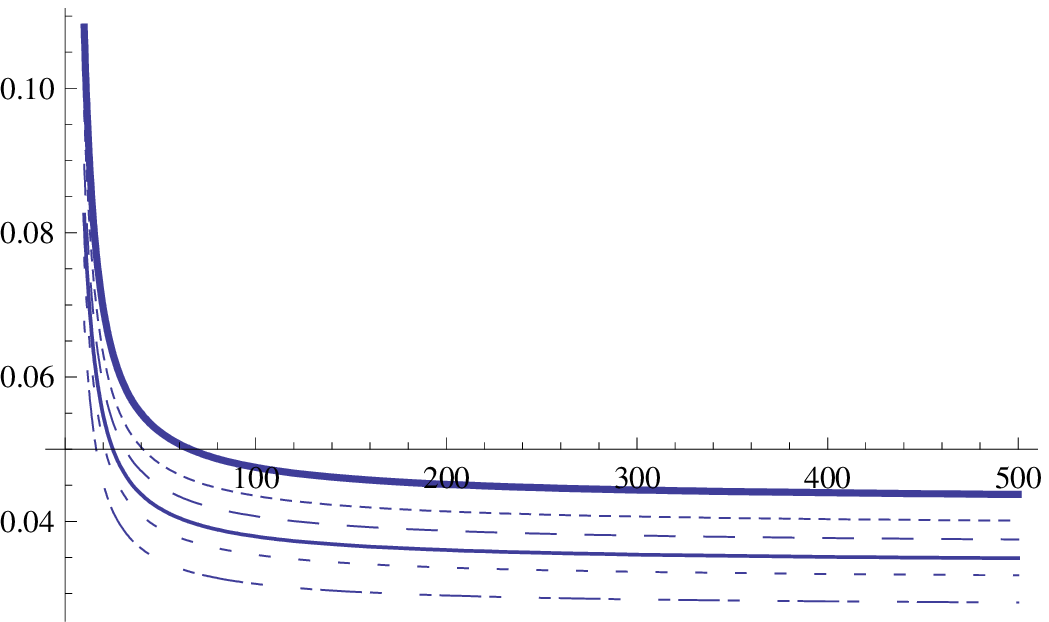}
}\\
{\hspace{-2.0cm}}
{\hspace{-2.0cm}}
{\hspace{5.0cm} $m_{\chi}\longrightarrow$  GeV}\\
{\hspace{-0.0cm} (a) \hspace{9.0cm} (b)}
 \caption{The same as in Fig. \ref{Fig:tr19} for the quantity $ {h_{coh}}$.}
 \label{Fig:th19}
   \end{center}
  \end{figure}

 Since the NFW profile preceded the VO profile and is more widely known, we are going to employ the velocity distribution obtained with this profile to compute the direct WIMP detection rates.
\\ We begin with the quantities $t$ and $h$, which are pretty independent of the particle model, but are sensitive to the WIMP velocity distribution and mass.
We first consider differential quantities  $dt/du$, which is proportional to the time average differential rate $dR/du$, and $dh/du$, which is  the relative differential modulated rate, as  functions of the energy transfer $Q$. For a typical light and a typical heavy target  are shown in Figs \ref{Fig:r127}-\ref{Fig:h19}. Two values of the WIMP mass, namely $30$ and $100$ GeV/$c^2$ were considered. Note that in some instances the differential modulated rate changes sign. This may result in cancellations in the integrated rate yielding smaller values than expected due to the smallness of $\delta$ alone. It may also cause a shift the maximum of the rate from June to December.
\\ By integrating the differential rate we obtain the total rates $t$ and $h$  as a function of the WIMP mass for a given energy threshold. Typical examples for the coherent process are shown in Figs \ref{Fig:tr127}-\ref{Fig:th19}, for a typical light target ($^{19}$F) and an intermediate-heavy target ($^{127}$I). Two typical threshold values  of zero and 10 keV were considered.
\section{Discussion}
\begin{figure}[t]
\begin{center}
\subfloat{
\rotatebox{90}{\hspace{0.0cm} {Event rate$\longrightarrow$ per kg-y
}}
\includegraphics[scale=0.6]{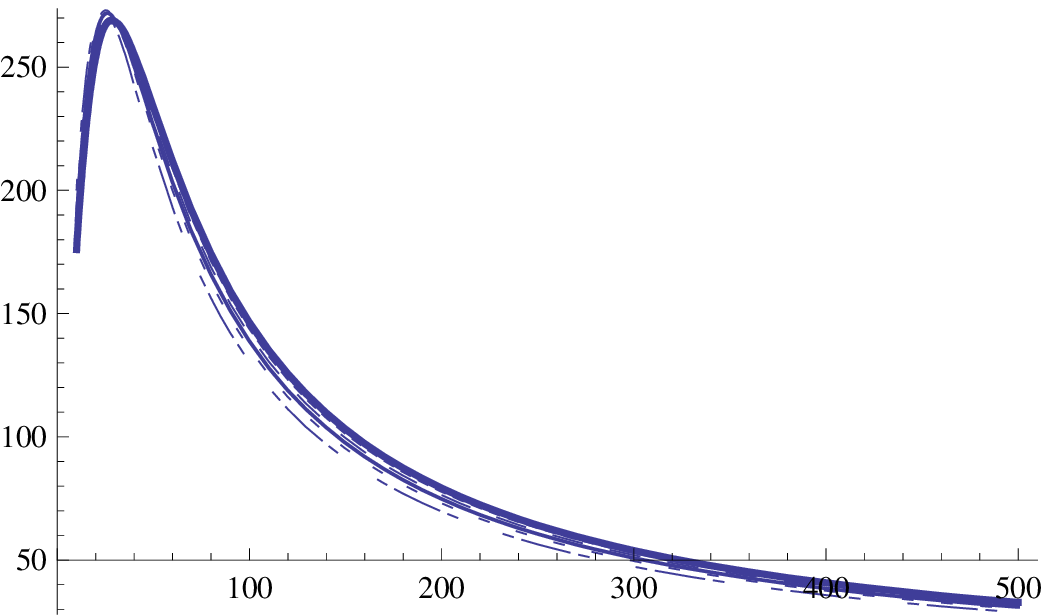}
}
\subfloat{
\rotatebox{90}{\hspace{0.0cm} {Event rate$\longrightarrow$ per kg-y
}}
\includegraphics[scale=0.6]{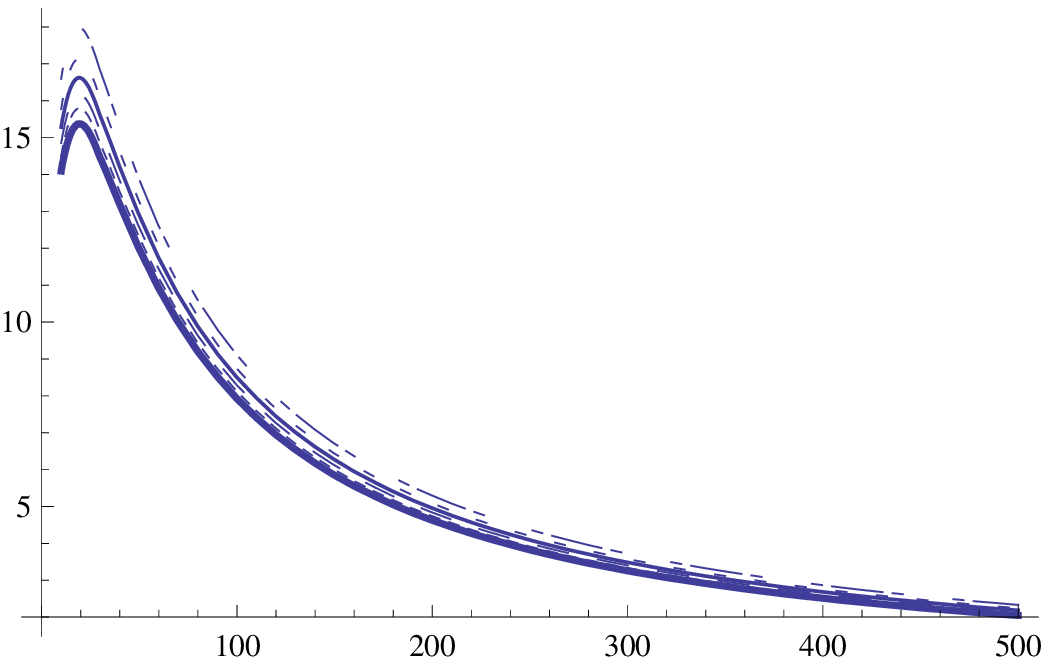}
}
{\hspace{5.0cm} $m_{\chi}\longrightarrow$  GeV}\\
{\hspace{-0.0cm} (a) \hspace{9.0cm} (b)}
\caption{The (time averaged) event rate per kg target per year as a function of the WIMP mass in GeV assuming a nucleon cross section $\sigma_N=10^{-6}$ pb. The results shown are due to the coherent mode for a heavy target (A=127) in (a) and a light target in (b) assuming zero threshold . The effect of the asymmetry is not clearly visible. At low WIMP masses the rate is suppressed due to small reduced mass.}
 \label{fig:Totrate}
 \end{center} 
  \end{figure}

\begin{figure}[t]
\begin{center}
\subfloat{
\rotatebox{90}{\hspace{0.0cm} {Event rate$\longrightarrow$ per kg-y}}
\includegraphics[scale=0.6]{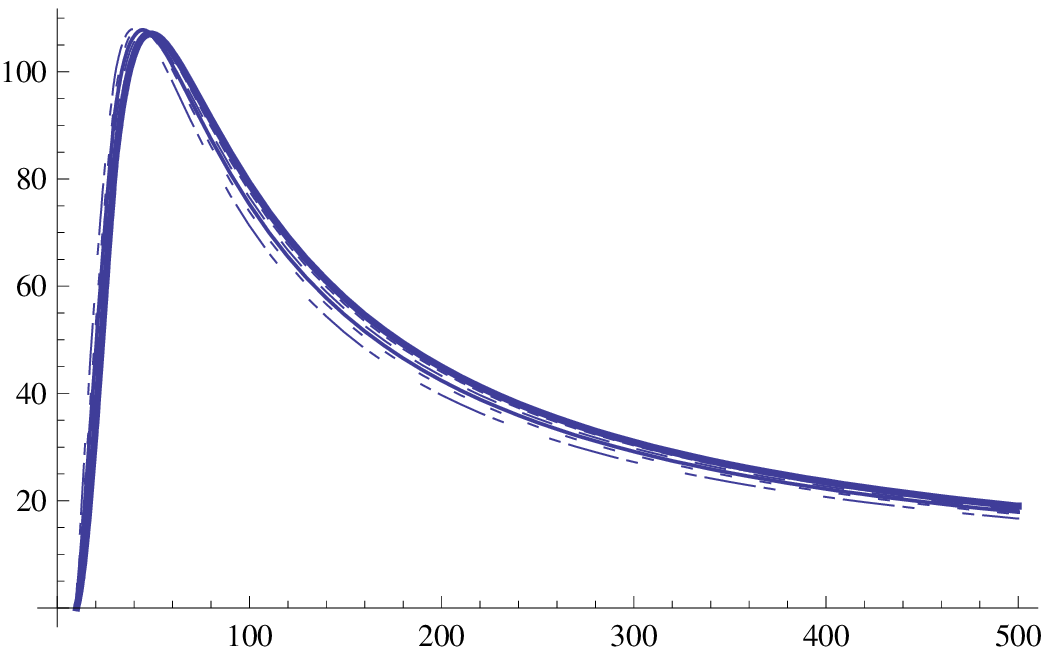}
}
\subfloat{
\rotatebox{90}{\hspace{0.0cm} {Event rate$\longrightarrow$ per kg-y}}
\includegraphics[scale=0.6]{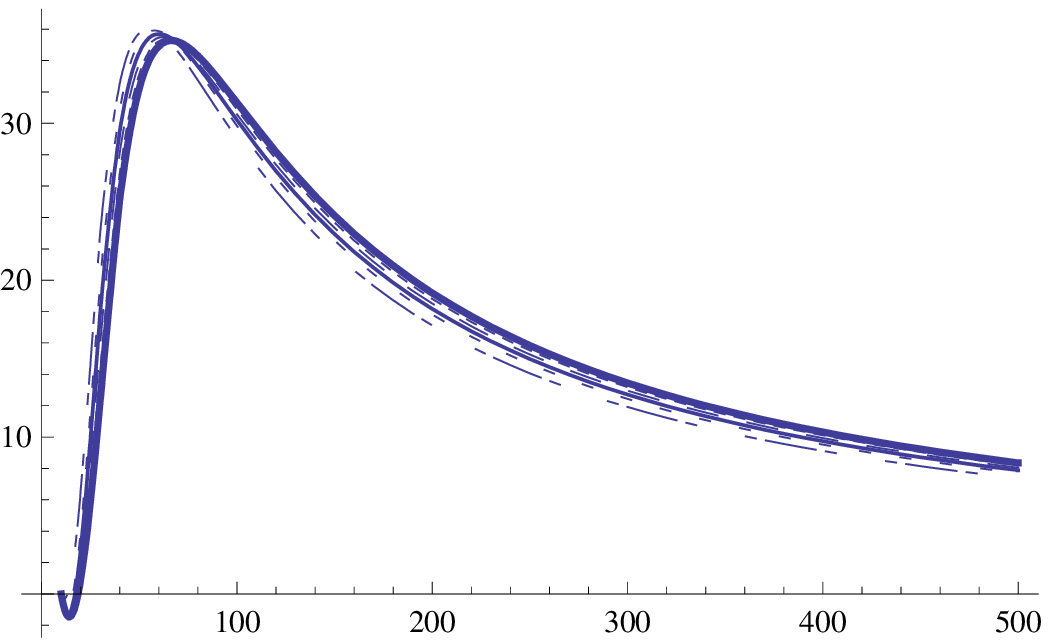}
}
{\hspace{5.0cm} $m_{\chi}\longrightarrow$  GeV}\\
{\hspace{-0.0cm} (a) \hspace{9.0cm} (b)}
\caption{The total rates in the case of a heavy target (A=127) computed assuming   an energy threshold of 10 keV without quenching (a) and with quenching (b) and  a nucleon cross section   $\sigma_N=10^{-6}$ pb. Note that the quenched rate is about a factor of 8 smaller compared to that at zero threshold  (see Fig. \ref{fig:Totrate}).}
 \label{fig:qTot127}
 \end{center} 
  \end{figure}

\begin{figure}[t]
\begin{center}
\subfloat{
\rotatebox{90}{\hspace{0.0cm} {Event rate$\longrightarrow$ per kg-y 
}}
\includegraphics[scale=0.6]{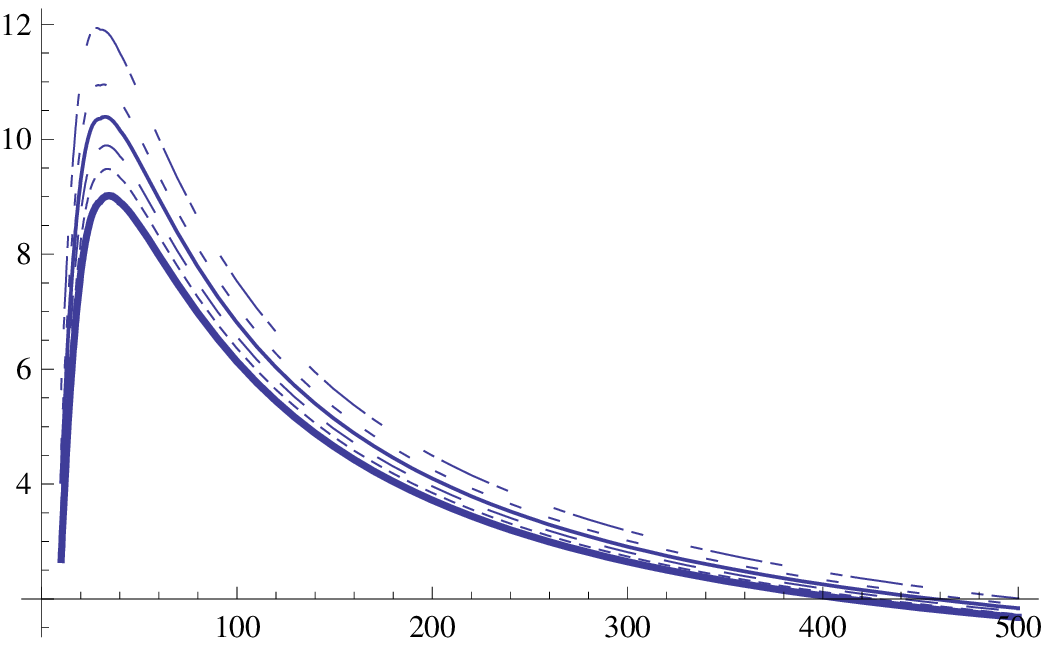}
}
\subfloat{
\rotatebox{90}{\hspace{0.0cm} {Event rate$\longrightarrow$ per kg-y
}}
\includegraphics[scale=0.6]{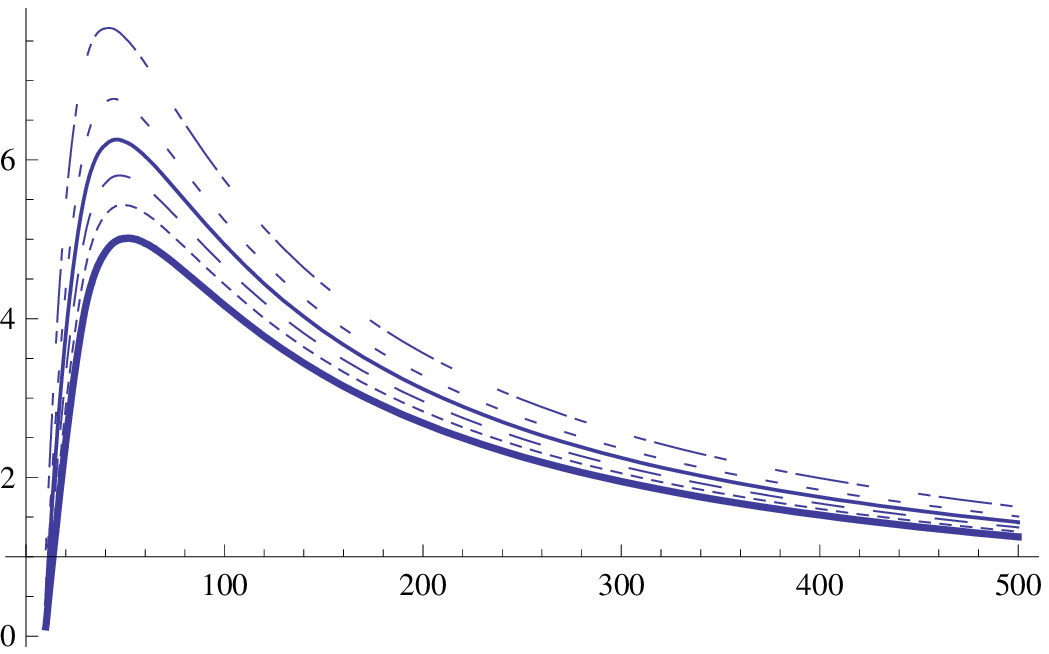}
}
{\hspace{5.0cm} $m_{\chi}\longrightarrow$  GeV}\\
{\hspace{-0.0cm} (a) \hspace{9.0cm} (b)}
\caption{The same as in Fig. \ref{fig:qTot127}  in the case of a light target (A=19). Here the quenched rate is about  a factor of two smaller than that at zero threshold (see Fig. \ref{fig:Totrate}).}
 \label{fig:qTot19}
 \end{center} 
  \end{figure}

With the above ingredients we can now compute the total event rate. We do not know what the elementary nucleon cross section is, but following the more or less standard practice,  we will assume it to be independent of the WIMP mass and equal to $10^{-6}$pb. We will also take the WIMP density in our vicinity to have the canonical value \cite{PDG} of 0.3 GeV/c$^2$ cm$^{-3}$.
 Employing Eq. (\ref{eventrate}) and using the values of $t$ discussed above  we find the results shown in Fig. \ref{fig:Totrate} for zero energy threshold. 
For an energy threshold value of 10 keV we obtain the results shown in Figs \ref{fig:qTot127} and \ref{fig:qTot19} for A=127 and A=19 respectively. On these 
figures we also show the effect of quenching \cite{VEREJ08}-\cite{FS08a} assuming an energy threshold of 10 keV. There is no need to show again  the effects of modulation on the rate, since the above shown results are adequate (h refers to the ratio of the modulated to the time averaged rate). 
   \section{Conclusions}
   In the present paper we first derived the WIMP velocity distribution for a spherically symmetric WIMP density profile. This was done in a self consistent way by applying Eddington's approach. Asymmetry of the velocity distribution was also taken into account by incorporating angular momentum into the expression of the energy entering the  phase space distribution function.
 Using this distribution function we obtained both the differential  and total rates entering direct WIMP detection. We find that the time averaged differential rates do not sensitively depend on the asymmetry parameter $\beta$. The obtained shape of this signal unfortunately  cannot really be differentiated from that expected from most backgrounds. The differential modulation rate $H$ shows some dependence on the asymmetry parameter, especially for light targets. We found that this differential rate changes sign at some energy transfer, which depends on the WIMP mass. Perhaps this signature may aid in discarding some season dependent backgrounds.\\
 After that we computed the corresponding total rates. We find that the time averaged total rates are not very sensitive to the asymmetry parameter, in agreement with earlier results obtained with distributions given in terms of
Tsallis type functions \cite{VerHanH}. The maximum rate is expected at a WIMP mass of about 75 GeV for a heavy target and 30 GeV for a light target. The modulation for a light system is always positive (maximum in June) and tends to increase with asymmetry. For zero threshold is  rising from $2h=4$ $\%$ to about $2h=6$ $\% $ as the asymmetry increases from $\beta=0$ to $\beta=0.5$. Assuming a threshold of 5 keV 
we get $2h=5-8$ $\%$. Higher modulations up to 16$\%$ can be expected at smaller WIMP masses. For a heavy target the modulation increases with asymmetry, but it goes through zero at a WMIP mass of about 50 GeV. We expect a positive  modulation at lower WIMP masses, with a maximum of about $2h=6\%$ and negative modulation in the case of heavier WIMPS ranging to  $2|h|=4$ to $6\%$ depending on the asymmetry parameter.

 Finally we have seen  that both the spherically symmetric and the axially symmetric velocity distributions obtained from the realistic NFW  density profile can be approximated by a Maxwell-Boltzmann velocity distribution in a finite domain, i.e. one in which the upper bound of the velocity (escape velocity) is not put in by hand but it comes naturally from the Eddington method. Investigations of the goodness of such a fit still continue. Anyway, as we have discussed in the section marked by DWER on its title, we expect the approximate solution to be appropriate in calculations relevant to dark matter searches.  So one may use this distribution in the future to simplify the calculations.

 We did not consider in this work directional experiments, i.e. experiments in which not only the energy but the direction of the recoiling nucleus is also observed \cite{DRIFT1},\cite{DRIFT2}. In such experiments both the observed time averaged rates as well as  the modulation are expected to be direction
   dependent \cite{VF07}. We expect that such experiments are going to be much more sensitive to the form of the velocity distribution and, in particular, the asymmetry parameter.
\section*{ Acknowledgments }
This work was  partly supported by the European Union under the contracts MRTN--CT--2004--503369 and
MRTN-CT-2006-035863 (UniverseNet). 

\end{document}